\documentclass[nofootinbib, reprint, preprintnumbers, amsmath, amssymb, amsfonts, aps, pra, superscriptaddress, floatfix]{revtex4-2}
\usepackage[utf8]{inputenc}
\usepackage{mathtools}
\usepackage{graphicx}
\usepackage{dcolumn}
\usepackage{bm}
\usepackage{physics}
\usepackage{float}
\usepackage{placeins}
\usepackage{color}
\usepackage{braket}
\usepackage[normalem]{ulem}
\usepackage{enumerate}
\usepackage{enumitem}
\usepackage[colorlinks=true,linkcolor=blue,citecolor=blue,urlcolor =blue]{hyperref}
\usepackage{mathrsfs}
\usepackage{bbm}
\usepackage{siunitx}
\usepackage{graphicx,color}
\usepackage{amssymb}
\usepackage{amsmath}
\usepackage{placeins} 
\usepackage{subcaption}
\usepackage{bm}
\usepackage{comment}
\usepackage{lipsum}
\newcommand{\YM}[1]{\textcolor[rgb]{0, 0, 0}{#1}}

\newcommand{\ym}[1]{\textcolor[rgb]{0, 0, 0}{#1}}
\newcommand{\tana}[1]{\textcolor[rgb]{0, 0, 0}{#1}}
\begin{document}
\title{
Theoretical Analysis of Photonic Resonances in Spectroscopic Measurements of a Kerr Nonlinear Resonator
}

\author{Yuki Tanaka}
\affiliation{Department of Electrical, Electronic, and Communication Engineering, Faculty of Science and Engineering, Chuo university, 1-13-27, Kasuga, Bunkyo-ku, Tokyo 112-8551, Japan}%
\affiliation{NEC-AIST Quantum Technology Cooperative Research Laboratory, National Institute of Advanced Industrial Science and Technology (AIST), Tsukuba, Ibaraki 305-8568, Japan}

\author{Aiko Yamaguchi}
 \affiliation{NEC-AIST Quantum Technology Cooperative Research Laboratory, National Institute of Advanced Industrial Science and Technology (AIST), Tsukuba, Ibaraki 305-8568, Japan}
 \affiliation{Secure System Platform Research Laboratories, NEC Corporation, 
1753, Shimonumabe, Kawasaki, Kanagawa 211-0011, Japan}

 \author{Tomohiro Yamaji}
 \affiliation{NEC-AIST Quantum Technology Cooperative Research Laboratory, National Institute of Advanced Industrial Science and Technology (AIST), Tsukuba, Ibaraki 305-8568, Japan}
 \affiliation{Secure System Platform Research Laboratories, NEC Corporation, 
1753, Shimonumabe, Kawasaki, Kanagawa 211-0011, Japan}

\author{Yuta Shingu}
\affiliation{Department of Physics, Graduate School of Science, Tokyo University of Science, Shinjuku, Tokyo 162-8601, Japan.}

\author{Keisuke Matsumoto}
\affiliation{Department of Physics, Graduate School of Science, Tokyo University of Science, Shinjuku, Tokyo 162-8601, Japan.}
  
\author{Tsuyoshi Yamamoto}
\affiliation{NEC-AIST Quantum Technology Cooperative Research Laboratory, National Institute of Advanced Industrial Science and Technology (AIST), Tsukuba, Ibaraki 305-8568, Japan}
\affiliation{Secure System Platform Research Laboratories, NEC Corporation, 
1753, Shimonumabe, Kawasaki, Kanagawa 211-0011, Japan}

\author{Yuichiro Matsuzaki}
\email{ymatsuzaki872@g.chuo-u.ac.jp}
\affiliation{Department of Electrical, Electronic, and Communication Engineering, Faculty of Science and Engineering, Chuo university, 1-13-27, Kasuga, Bunkyo-ku, Tokyo 112-8551, Japan}%

\date{\today}


\begin{abstract}
The Kerr parametric oscillator (KPO) has recently attracted considerable attention from the perspective of its applications to quantum information processing, and understanding its properties is an important challenge. Spectroscopic measurements serve as an effective means of elucidating detailed information about the system, such as the energy-level structure and the transition matrix elements of the KPO. Conventional spectroscopy requires the drive frequency to match an energy spacing with a nonzero transition matrix element. In recent years, a phenomenon called photonic resonance (PR) has been theoretically predicted in KPO spectroscopy. Specifically, resonance occurs under the condition that the detuning is set to $n/2$ times the Kerr nonlinearity, where $n$ is a natural number. However, under this condition the transition matrix element vanishes, and thus the mechanism by which photonic resonance (PR) arises has remained unclear. In this work, we aim to elucidate the physical origin of PR observed in KPO spectroscopy. We first performed theoretical calculations and experiments of spectroscopic measurements, confirming that PR can indeed be observed and that the theoretical and experimental results are in qualitative agreement. We then carried out an analytical study under the assumption of an ideal noiseless environment. Our analysis revealed that, although the transition matrix element of the external field expressed in the system’s energy eigenbasis is zero, higher-order perturbative effects induce Rabi oscillations between the ground and excited states. Furthermore, numerical simulations in a time domain including the effect of decoherence demonstrated that coherent oscillations decay, leading to the appearance of PR.
\end{abstract}

\maketitle
\section{Introduction}
\label{sec:intro}
In recent years, parametric oscillators with Kerr nonlinearity (Kerr parametric oscillators, KPOs) \cite{cochrane1999macroscopically,goto2016bifurcation}, realized by parametrically driving in superconducting resonators \cite{bourassa2012josephson,meaney2014quantum,leghtas2015confining}, have attracted significant attention in the field of quantum 
\textcolor{black}{technology}
\cite{goto2019demand,wang2019quantum,yamaji2022spectroscopic}. 
\ym{A KPO can be viewed as the quantum analogue of the classical parametron \cite{goto1959parametron}. To exhibit genuine quantum behavior, its Kerr nonlinearity must exceed the photon-loss rate \cite{kirchmair2013observation,milburn1991quantum,wielinga1993quantum,goto2016bifurcation}.}
\textcolor{black}{The KPO is also a promising platform for investigating fundamental quantum phenomena, including phase transitions \cite{wang2020excited,chavez2023spectral,garcia2025impact} and quantum tunneling through the barrier of a double-well potential \cite{reynoso2023quantum,venkatraman2022driven,frattini2024observation,venkatraman2024nonlinear}.}

The two stable phase states of a KPO can be utilized as logical states of a qubit \cite{cochrane1999macroscopically,goto2016bifurcation,grimm2020stabilization,mirrahimi2014dynamically,hajr2024high}. Accordingly, KPOs have been proposed as promising platforms for quantum annealing \cite{goto2016bifurcation,puri2017quantum,nigg2017robust,kanao2021high} as well as for fault-tolerant universal quantum computation \cite{cochrane1999macroscopically,goto2016universal,puri2017engineering,puri2020bias}. In addition, KPOs are being explored for various other applications in quantum information processing, such as fast and high-fidelity qubit readout \cite{lin2014josephson,yamamoto2016parametric,suzuki2023quantum,krantz2016single} and single-qubit operations \cite{grimm2020stabilization}.

Clarifying the physical properties of \tana{KPOs} under parametric driving—particularly its energy-level structure—is of central importance from both fundamental and applied perspectives \cite{masuda2021theoretical}. Spectroscopy provides a powerful and precise means to characterize the energy spectrum. 
\ym{Previous studies have characterized the energy-level structure of KPOs, as well as its drive-strength dependence, via pulse-excitation spectroscopy and continuous-wave reflection spectroscopy}
\cite{yamaji2022spectroscopic,yamaguchi2024spectroscopy}. 
In general, resonance in spectroscopy requires two conditions to be satisfied: (1) \tana{the angular frequency $\omega$ of the external drive corresponds to the energy level difference of the system divided by Planck’s constant $\hbar$}, and (2) the transition matrix element of the drive Hamiltonian, expressed in the energy eigenbasis, must be 
\ym{non-zero}.

Recently, however, a phenomenon called photonic resonance (PR), which does not follow the conventional resonance conditions, has been theoretically predicted in KPO spectroscopy \cite{bartolo2016exact,roberts2020driven} \tana{, and has also been observed experimentally\cite{yamaji2023correlated}}. Specifically, resonance is predicted to occur when the detuning is set to $n/2$ times the Kerr nonlinearity coefficient, where $n$ is a natural number. Here, detuning is defined as the difference between half of the two-photon drive frequency and the resonance frequency of the KPO. 
\ym{Under this condition, the transition matrix elements vanish, so the standard resonance mechanism cannot account for the phenomenon, and its physical origin remains unclear. To the best of our knowledge, no explanation for this issue has been provided in the previous literature.}

In this work, we aim to elucidate the physical origin of PR through a combined experimental and theoretical approach. First, we performed theoretical calculations of spectroscopic response and corresponding experiments, confirming that PR can indeed be observed and that theory and experiment show qualitative agreement. 
\ym{Furthermore, under idealized conditions without decoherence, analytical calculations to \tana{the} third order in perturbation theory show that higher-order terms induce Rabi oscillations between the ground and excited states. This occurs even though the leading transition matrix elements vanish.}
Finally, numerical simulations \ym{in a time domain} including decoherence demonstrated that the coherent oscillations decay over time, resulting in the emergence of PR.
\textcolor{black}{PR proceeds in two stages: a dominant stage of higher-order–induced Rabi oscillations followed by their damping due to decoherence.}

\section{KPO Hamiltonian and Resonance Conditions}
We begin by describing the Kerr nonlinear resonator.
Throughout this paper, we set $\hbar = 1$.
The Hamiltonian of a Kerr nonlinear resonator in the laboratory frame is defined as

\begin{equation}
H_{\text{lab}}(t) = \frac{\chi}{2} a^{\dagger} a^{\dagger} a a + \omega_c a^{\dagger} a + 2p \left( a^2 + a^{\dagger 2} \right)\cos \omega_p t,
\label{eq:H_lab}
\end{equation}
where $\omega_c$ is the resonance frequency of the Kerr nonlinear resonator, $\chi$ is the Kerr nonlinearity coefficient, $p$ is the amplitude of the parametric drive, and $\omega_p$ is the drive frequency.
By moving to the rotating frame and applying the rotating-wave approximation, Eq.~\eqref{eq:H_lab} can be rewritten as
\begin{equation}
H = \frac{\chi}{2} a^{\dagger} a^{\dagger} a a + \Delta a^{\dagger} a
+ p \left( a^{2} + a^{\dagger 2} \right),
\label{h_kpo}
\end{equation}
where $\Delta=\omega_c-\omega_p/2$ is the detuning.
\textcolor{black}{Is is known that the ground states of this Hamiltonian are degenerate~\cite{cochrane1999macroscopically,goto2016bifurcation}.}
Defining
\begin{align}
H_0 = \frac{\chi}{2} a^{\dagger} a^{\dagger} a a + \Delta a^{\dagger} a, \
V = a^{2} + a^{\dagger 2},
\end{align}
Eq.~\eqref{h_kpo} can be expressed as
\begin{equation}
H = H_0 + p V.
\end{equation}
Now, considering the case $p=0$, $H_0$ is diagonal, and the eigenvalues and eigenstates are given by
\begin{equation}
H_0|n\rangle = E_n ^{(0)}|n\rangle, \quad
E_n ^{(0)}= \Delta n + \frac{\chi}{2}n(n-1).
\end{equation}
When $E_0^{(0)} = E_n^{(0)}$, the ground state ($|0\rangle$) and the excited state ($|n\rangle$) become degenerate. The degeneracy condition is therefore
\begin{equation}
\Delta = - \frac{\chi}{2}(n-1).
\label{detu}
\end{equation}
\textcolor{black}{It is known that this condition is essential to control the interference of the tunneling effect between ground states of the KPO\cite{venkatraman2022driven}. Also,}
previous studies have shown that, under 
the Eq. \eqref{detu}, a resonance phenomenon called photonic resonance (PR) occurs \cite{bartolo2016exact,roberts2020driven}. However, for $n>2$, one finds
$\langle n|V|0\rangle = 0$
meaning that no direct transition exists from $|n=0\rangle$ to $|n\rangle$. Consequently, the mechanism by which PR arises under these conditions has not been fully understood.

\section{
Comparison between experimental and theoretical results
}

\begin{figure*}[h!]
    \centering

    \begin{subfigure}[t]{0.45\linewidth}
        \centering
        \includegraphics[width=\linewidth]{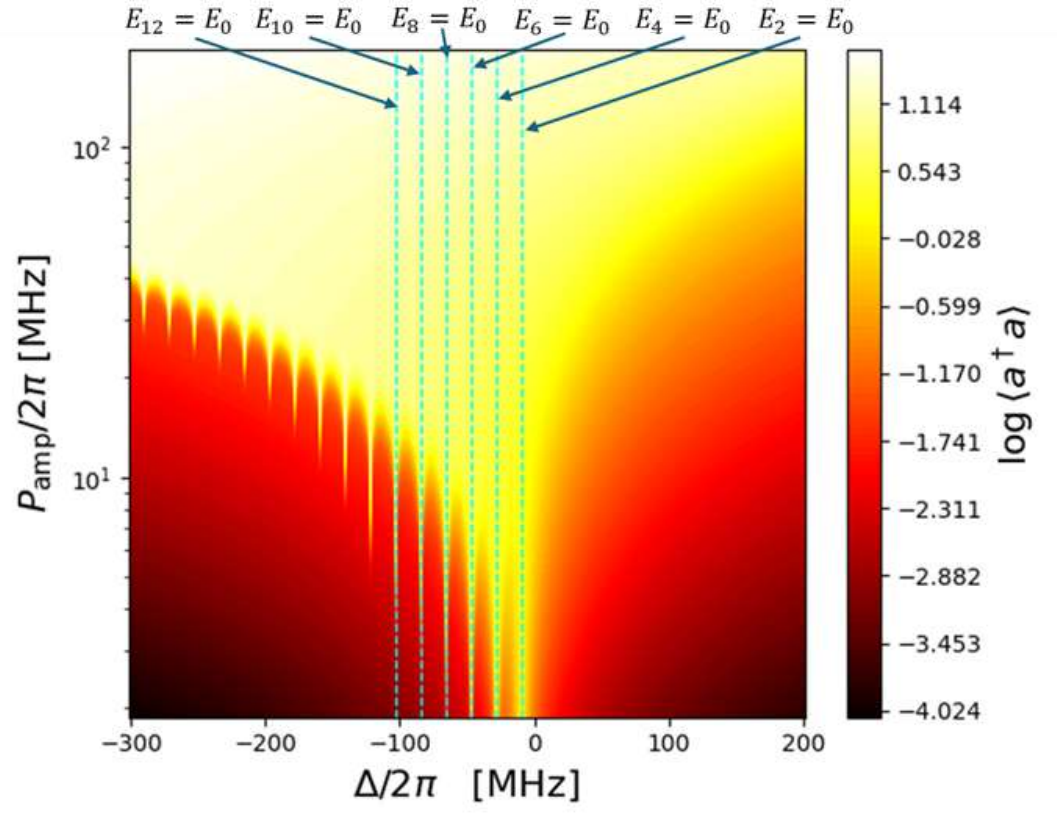}
        \caption{Analytical result of the steady-state photon number of the KPO. 
        The $x$-axis shows the detuning $\Delta/2\pi\,[\mathrm{MHz}]$, the $y$-axis the parametric drive $p/2\pi\,[\mathrm{MHz}]$, and the $z$-axis the photon number $\log_{10}\langle a^{\dagger}a\rangle$. 
        The dotted lines indicate the detuning values that satisfy the degeneracy condition 
        $\Delta = -\frac{\chi}{2}(n-1)$ for $n=2,4,6,8,10,12$. 
        The parameters are set to a dissipation rate $\kappa /2\pi =10^{-6}$ MHz and a nonlinearity $\chi/2\pi = -18.729\,\mathrm{MHz}$.
        }
        \label{bart}
    \end{subfigure}
    \hfill
    \begin{subfigure}[t]{0.45\linewidth}
        \centering
        \includegraphics[width=\linewidth]{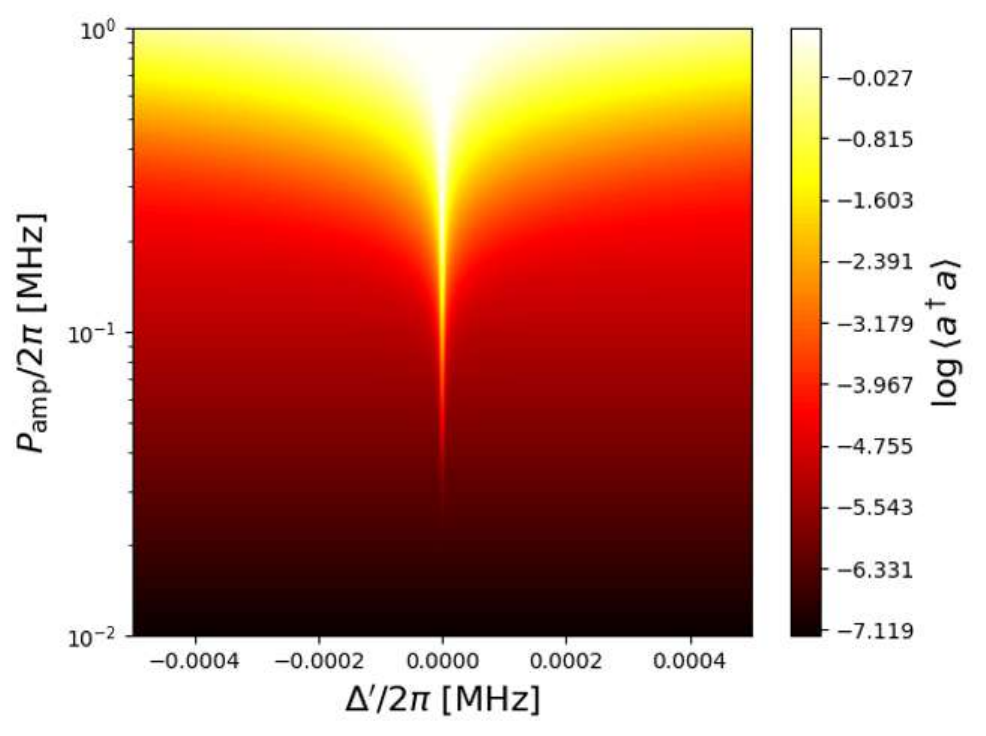}
        \caption{Enlarged analytical plot of the steady-state photon number of the KPO near the degeneracy between $|0\rangle$ and $|8\rangle$. 
        The $x$-axis shows the deviation $\Delta'/2\pi$ from the degeneracy condition at $-65.5515$ MHz, while the $y$-axis and $z$-axis are the same as in (a).
        }
        \label{sh08}
    \end{subfigure}

    \vspace{1em}
    
    \begin{subfigure}[t]{0.45\linewidth}
       \centering
       \includegraphics[width=\linewidth]{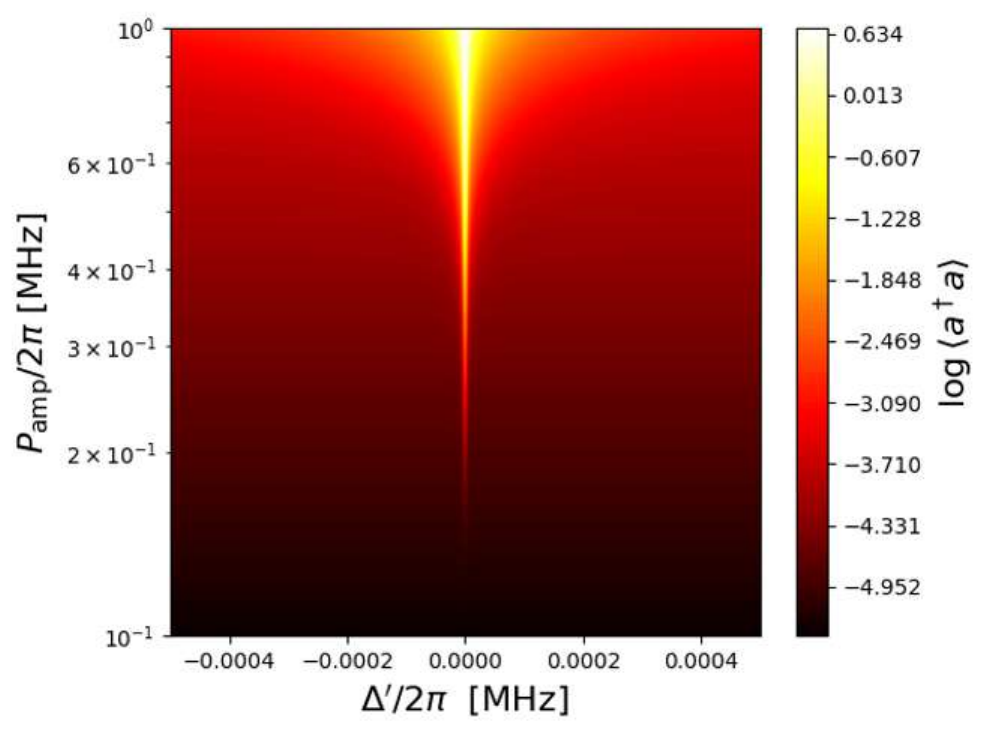}
       \caption{Enlarged analytical plot of the steady-state photon number of the KPO near the degeneracy between $|0\rangle$ and $|10\rangle$. 
       The $x$-axis shows the deviation $\Delta'/2\pi$ from the degeneracy condition at $-84.2805$ MHz, while the $y$-axis and $z$-axis are the same as in (a).
       }
       \label{sh10}
    \end{subfigure}
    \hfill
    \begin{subfigure}[t]{0.45\linewidth}
       \centering
       \includegraphics[width=\linewidth]{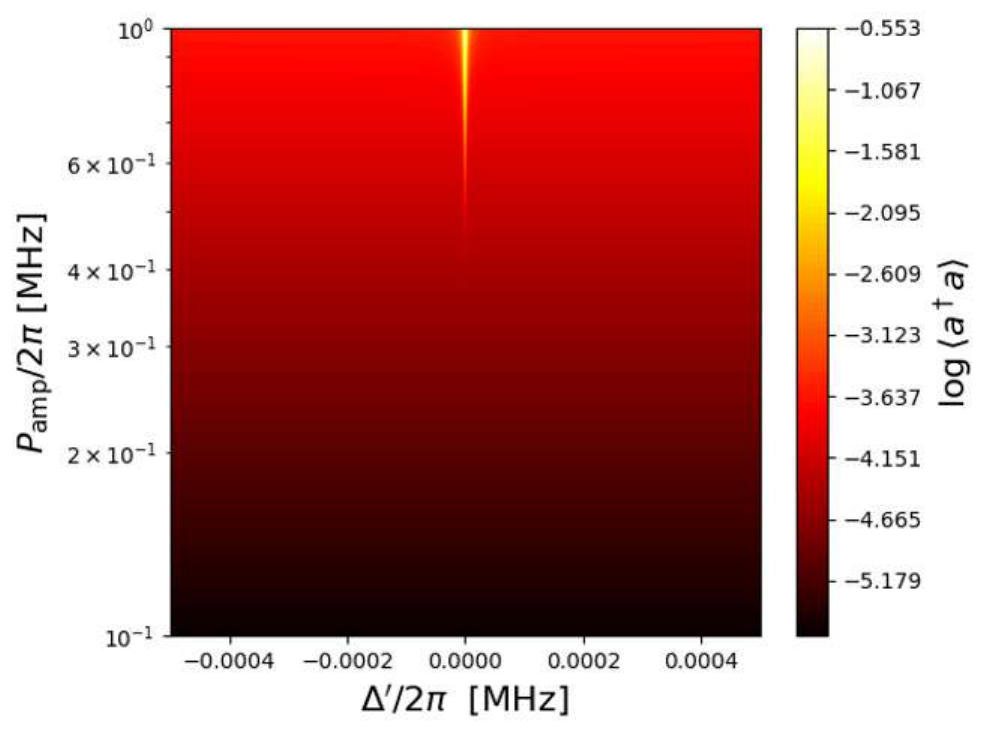}
       \caption{Enlarged analytical plot of the steady-state photon number of the KPO near the degeneracy between $|0\rangle$ and $|12\rangle$. 
       The $x$-axis shows the deviation $\Delta'/2\pi$ from the degeneracy condition at $-103.0095$ MHz, while the $y$-axis and $z$-axis are the same as in (a).
       }
        \label{sh12}
    \end{subfigure}

    \caption{Analytical plots of the steady-state photon number $\log_{10}\langle a^{\dagger}a\rangle$ of the KPO. 
    The parameters are set to dissipation rate $\kappa/2\pi = 10^{-6}\,\mathrm{MHz}$ and nonlinearity $\chi/2\pi = -18.729\,\mathrm{MHz}$.}
    \label{barall}
\end{figure*}

\begin{figure*}[t]
    \centering
    \begin{subfigure}[t]{0.45\linewidth}
       \centering
       \includegraphics[width =\linewidth]{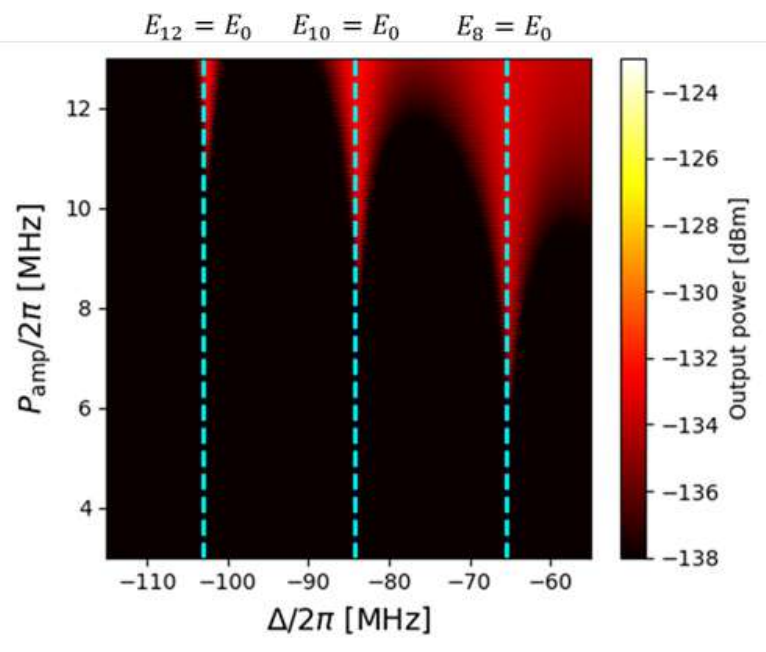}
       \caption{
       Analytical result of the steady-state photon number of the KPO. 
       The $x$-axis shows the detuning $\Delta/2\pi\,[\mathrm{MHz}]$, the $y$-axis the parametric drive $p/2\pi\,[\mathrm{MHz}]$, and the $z$-axis the output power. 
       The plots are enlarged near the degeneracy points $E_{12}=E_0$, $E_{10}=E_0$, and $E_8=E_0$.
       }
        \label{bar_zoom}
    \end{subfigure}
    \hfill
    \begin{subfigure}[t]{0.45\linewidth}
       \centering
       \includegraphics[width=\linewidth]{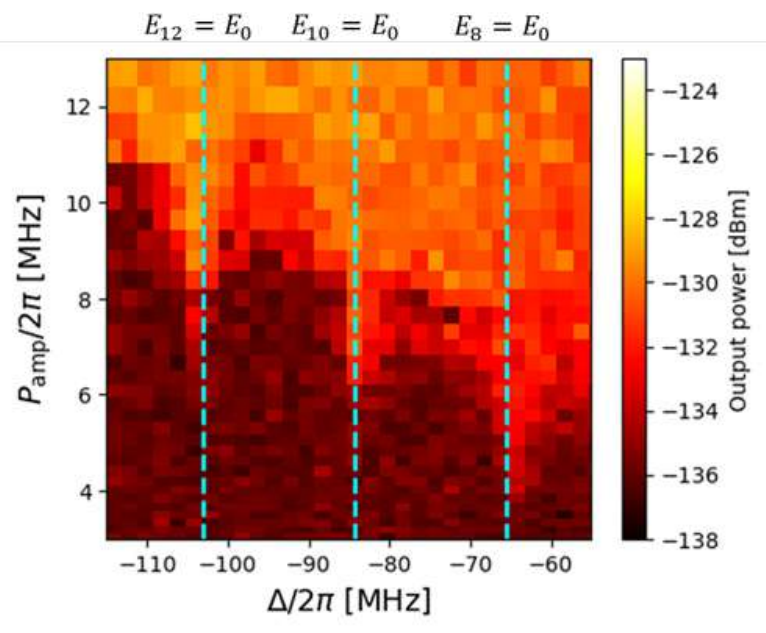}
       \caption{Experimental result of spectroscopic measurements on the KPO. 
       The $x$-, $y$-, and $z$-axes are the same as in (a).
       }
        \label{exper_zoom}
    \end{subfigure}
    \caption{Comparison between analytical and experimental results of the output power of the KPO. 
    For the numerical calculations, we adopt experimental parameters: external dissipation rate $\kappa_e/2\pi=0.47$ MHz, internal dissipation rate $\kappa_i/2\pi=0.26$ MHz, total dissipation rate $\kappa/2\pi = (\kappa_e +\kappa_i)/2\pi = 0.73\,\mathrm{MHz}$, and nonlinearity $\chi/2\pi = -18.729\,\mathrm{MHz}$.}
    \label{fig:placeholder}
\end{figure*}

In this section, we compare the analytically derived steady-state photon number with experimental spectroscopic results and confirm that PR can indeed be observed.  
For the theoretical analysis, we employ the GKSL (Gorini-Kossakowski-Sudarshan-Lindblad) master equation to describe the time evolution of the system including decoherence, using the Hamiltonian given by Eq.~\eqref{h_kpo}.  
Assuming single-photon loss as the noise source, the master equation reads
\begin{equation}
\frac{d\rho}{dt} = -i[H, \rho] + \frac{\kappa}{2}\left(2a\rho a^{\dagger}-\{a^{\dagger}a,\rho\}\right),
\label{GKSL}
\end{equation}
where $\kappa=\kappa_e+\kappa_i$ denotes the total dissipation rate, with $\kappa_e$ and $\kappa_i$ being the external and internal dissipation rates, respectively.  
The steady-state photon number of this master equation has been analytically derived by Bartolo \cite{bartolo2016exact}.  
Figures~\ref{bart}--\ref{sh12} show the analytical results of $\log_{10}\langle a^{\dagger}a\rangle$ as functions of detuning and parametric drive, with parameters $\chi/2\pi = -18.729$ MHz and $\kappa /2\pi =10^{-6}$ MHz.  
Resonances are observed around 1 MHz parametric drive between $|0\rangle$ and $|2\rangle$, $|0\rangle$ and $|4\rangle$, and $|0\rangle$ and $|6\rangle$.  
Similarly, even for \tana{$|0\rangle$ and $|8\rangle$, $|0\rangle$ and $|10\rangle$, and $|0\rangle$ and $|12\rangle$}, resonances are visible around 1 MHz when focusing on the enlarged regions.  

For comparison with experiment, we use the analytical solution with dissipation parameters matching the experimental values and plot the results in Fig.~\ref{bar_zoom}.  
Here, the analytical photon number is converted into the output power $P_o$ 
\tana{[W]} using
$P_o = \hbar \omega_r \kappa_e \langle a^{\dagger}a\rangle$
allowing direct comparison with the experimental data shown in Fig.~\ref{exper_zoom}.  

\begin{figure}[t]
    \centering
    \includegraphics[width =7.5cm]{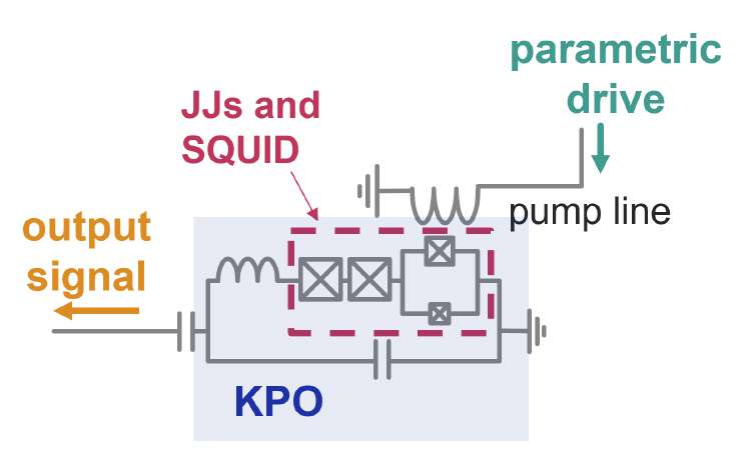}
    \caption{
       \textcolor{black}{
       Schematic circuit diagram of the KPO.
       The parametric pump is injected from the pump line, which is inductively coupled to the SQUID in the KPO.
       When the internal photon number increases due to PR and related effects, it is detected as an output signal.
       The output signal is amplified by amplifiers located both inside and outside the refrigerator (not shown) and then measured using a spectrum analyzer.
       } 
       }
    \label{exp_diagram}
\end{figure}

\textcolor{black}{
In the experiment, a Josephson parametric oscillator (JPO) incorporating a Josephson junction and a SQUID was used as the KPO~\cite{yamaguchi2024spectroscopy}. 
The KPO was placed in a dilution refrigerator and cooled down to approximately 10 mK. 
When a parametric drive was applied through the pump line magnetically coupled to the SQUID, the output signal from the KPO was measured using a spectrum analyzer, allowing us to observe photonic resonance (PR) (see Fig.~\ref{exp_diagram}). 
\textcolor{black}{
The horizontal axis represents the detuning, while the vertical axis represents the amplitude of the parametric drive.
}
}

\textcolor{black}{
A comparison between the analytical results and the experimental data shows that, in both cases, the photon number increases sharply at the degeneracy condition $\Delta = -\frac{\chi}{2}(n-1)$, confirming that PR is indeed observed experimentally. 
However, the parametric drive amplitude $P_{\rm amp}$ at which PR starts to appear does not perfectly coincide between experiment and theory; the theoretical value is approximately 1.5 times higher than the experimental one. 
The parametric drive signal, generated by a room-temperature signal generator, is delivered to the KPO pump line inside the dilution refrigerator through multiple components such as cables, attenuators, and filters, which cause power attenuation. 
Therefore, when estimating $P_{\rm amp}$ based on the line attenuation, an uncertainty of about 2~dB \tana{can remain}. 
We attribute the slight discrepancy between the theoretical and experimental thresholds of $P_{\rm amp}$ for the onset of PR to this uncertainty.
}

\section{Rabi oscillations between the ground and excited states induced by higher-order perturbative effects}
We explain the mechanism by which Rabi oscillations occur between the ground and excited states under the degeneracy condition
$\Delta n + \frac{\chi}{2}n(n-1)=0$.
As mentioned earlier, for $n>2$, the transition matrix element of the drive Hamiltonian,
\(\langle n|V|0\rangle\),
vanishes under this condition. 
\ym{So the standard resonance mechanism cannot account for the phenomenon of PR, and its physical origin remains unclear. To the best of our knowledge, no explanation for this issue has been provided in the previous literature.}
However, by analyzing higher-order terms using perturbation theory in combination with numerical calculations, we demonstrate in this section that Rabi oscillations are nevertheless induced between the ground and excited states.

The Hamiltonian is given by
$H = H_0 + pV$,
While $H_0$ can be analytically diagonalized, in general, the analytical diagonalization of $H$ is difficult. Assuming $p$ to be sufficiently small, we apply perturbation theory. The eigenstates of $H_0$ are Fock states, and under the degeneracy condition $\Delta n + \tfrac{\chi}{2}n(n-1)=0$, two Fock states become degenerate, which we denote as $|n_1\rangle$ and $|n_2\rangle$.

The Schrödinger equation for the eigenvalue problem is
\begin{equation}
    H|\psi_n\rangle =E_n |\psi_n\rangle,
    \label{ham_e}
\end{equation}
and we expand both sides perturbatively as
\begin{equation}
    H|\psi_n\rangle =(H_0 + pV)(|\phi\rangle +p|\psi_n^{(1)}\rangle +p^2|\psi_n^{(2)}\rangle +p^3|\psi_n^{(3)}\rangle + \cdots ),
    \label{degene_ham}
\end{equation}
\begin{equation}
\begin{split}
    E_n|\psi_n\rangle =&(E_n^{(0)}+ pE_n^{(1)}+ p^2E_n^{(2)}+ p^3E_n^{(3)}+ \cdots)
    \\ &(|\phi\rangle +p|\psi_n^{(1)}\rangle +p^2|\psi_n^{(2)}\rangle +p^3|\psi_n^{(3)}\rangle + \cdots ),
    \label{degene_e}
\end{split}
\end{equation}
where $|\phi\rangle$ is a linear combination of the degenerate eigenstates of the unperturbed Hamiltonian,
\begin{equation}
|\phi\rangle = c_1|n_1\rangle + c_2|n_2\rangle,
\end{equation}
with $c_1, c_2\in \mathbb{C}$. We impose the condition
\begin{equation}
\langle \phi|\psi_n^{(k)}\rangle =0 \quad (k\geq 1).
\end{equation}
By collecting terms of the same order in $p$, the zeroth-order contribution yields
\begin{equation}
H_0|\phi\rangle = E_n^{(0)}|\phi\rangle.
\end{equation}
The first-order terms give
\begin{equation}
 (H_0 -E_n^{(0)})|\psi_n^{(1)}\rangle = (E_n^{(1)}-V)|\phi\rangle,
     \label{p1}
\end{equation}
which serves as the starting point for degenerate perturbation theory.

\subsection{Perturbation calculation for the $|0\rangle,|2\rangle$ degeneracy}

We now consider the perturbative treatment of the case in which $|0\rangle$ and $|2\rangle$ are degenerate in $H_0$. In this situation, it is in fact possible to analytically diagonalize the total Hamiltonian, and the transition matrix element $\langle 2|V|0\rangle$ takes a finite value. However, for higher Fock states \ym{where $|0\rangle$ and $|n\rangle$ are degenerate for $n>2$ in $H_0$}, analytic diagonalization becomes 
\ym{difficult}. Therefore, in preparation for such cases, we present here the perturbative procedure using the $|0\rangle$ and $|2\rangle$ degeneracy as an illustrative example.  

From Eq.~\eqref{detu}, the degeneracy condition gives $\Delta=-\chi/2$. Multiplying Eq.~\eqref{p1} by $\langle n_1|$, and noting that $H_0|n_1\rangle=E_n^{(0)}|n_1\rangle$, we obtain
\begin{gather}
  \langle n_1|(H_0 -E_n^{(0)})|\phi_n^{(1)}\rangle = \langle n_1|(E_n^{(1)}-V)|\phi\rangle \notag \\
  c_1(\langle n_1|V|n_2\rangle-E_n^{(1)})+c_2\langle n_1|V|n_2\rangle =0.
  \label{1jin1}
\end{gather}
Similarly, multiplying Eq.~\eqref{p1} by $\langle n_2|$ and using $H_0|n_2\rangle=E_n^{(0)}|n_2\rangle$, we find
\begin{equation}
    c_1\langle n_2|V|n_1\rangle + c_2(\langle n_2|V|n_2\rangle-E_n^{(1)})=0.
    \label{1jin2}
\end{equation}
Equations \eqref{1jin1} and \eqref{1jin2} can be expressed in matrix form as
\begin{equation}
\left(\begin{array}{cc}
\left\langle n_1\right| V\left|n_1\right\rangle-E_{n}^{(1)} & \left\langle n_1\right| V\left|n_2\right\rangle \\
\left\langle n_2\right| V\left|n_1\right\rangle & \left\langle n_2\right| V\left|n_2\right\rangle-E_n^{(1)}
\end{array}\right) \cdot\binom{c_1}{c_2}=\binom{0}{0}.
\label{mat1ji}
\end{equation}
The secular equation is therefore given by
\begin{equation}
\det \begin{pmatrix}
\langle n_1 | V | n_1 \rangle - E_n^{(1)} & \langle n_1 | V | n_2 \rangle \\
\langle n_2 | V | n_1 \rangle & \langle n_2 | V | n_2 \rangle - E_n^{(1)}
\end{pmatrix} = 0.
\label{ei1ji}
\end{equation}

We 
\ym{consider a special case of}
$|n_1\rangle=|0\rangle$ and $|n_2\rangle=|2\rangle$. The relevant matrix elements are evaluated as
\begin{align*}
\langle n_1|V|n_1\rangle &= \langle 0|(a^2 +a^{\dagger 2})|0\rangle =0, \\
\langle n_2|V|n_2\rangle &= \langle 2|(a^2 +a^{\dagger 2})|2\rangle =0, \\
\langle n_1|V|n_2\rangle &= \langle 0|(a^2 +a^{\dagger 2})|2\rangle =\sqrt{2}, \\
\langle n_2|V|n_1\rangle &= \sqrt{2}.
\end{align*}
Substituting these results into Eq.~\eqref{ei1ji}, we obtain
\begin{equation}
E_n^{(1)}=\pm \sqrt{2}.
\end{equation}
Since $H_0|0\rangle =0|0\rangle$ and $H_0|2\rangle =0|2\rangle$, we have $E_n^{(0)}=0$. Thus, the energy eigenvalues up to first order are
\begin{equation}
    E_{\pm}=E_n^{(0)}\pm pE_n^{(1)} =\pm\sqrt{2}p.
\end{equation}

Next, we consider the corresponding eigenstates. Let the eigenstates associated with $E_{\pm}$ be denoted as $|E_{\pm}\rangle$. For $E_n^{(1)}=+\sqrt{2}$, Eq.~\eqref{mat1ji} yields $c_1=c_2$. After normalization, we obtain $c_1=c_2=1/\sqrt{2}$, and hence
\begin{equation}
|E_+\rangle=\frac{1}{\sqrt{2}}(|0\rangle+|2\rangle).
\end{equation}
Similarly, for $E_n^{(1)}=-\sqrt{2}$, the eigenstate is
\begin{equation}
|E_-\rangle=\frac{1}{\sqrt{2}}(|0\rangle-|2\rangle).
\end{equation}

Assuming the initial state is
\begin{equation}
|\psi(0)\rangle =|0\rangle=\frac{1}{\sqrt{2}}(|E_+\rangle+|E_-\rangle),
\end{equation}
its time evolution is given by
\begin{align}
|\psi(t)\rangle &= e^{-iHt}|\psi(0)\rangle \notag \\
&\simeq \frac{1}{\sqrt{2}} e^{-i E_+ t } |E_+\rangle + \frac{1}{\sqrt{2}} e^{-i E_- t } |E_-\rangle \notag \\
&= \cos\left( \sqrt{2}p t \right) |0\rangle
- i \sin\left( \sqrt{2}p t \right) |2\rangle.
\end{align}
Therefore, the probability of finding the system in $|2\rangle$ is
\begin{equation}
P_{0\rightarrow 2}(t) = |\langle 2|\psi(t)\rangle|^2 =\sin ^2(\sqrt{2}p t )= \frac{1}{2} \left( 1 - \cos\left( 2\sqrt{2}pt \right) \right).
\end{equation}

In conclusion, we confirm that Rabi oscillations occur between $|0\rangle$ and $|2\rangle$.

\subsection{Perturbation calculation for the $|0\rangle, |4\rangle$ degeneracy}

Proceeding to second-order perturbation theory, we obtain the effective 
\ym{driving} between $|0\rangle$ and $|4\rangle$. The transition probability becomes
\begin{equation}
   P_{0\rightarrow 4}(t) = |\langle 4|e^{-iHt}|0\rangle|^2 =\frac{1}{2} \left(1-\cos\frac{2\sqrt{6}p^2}{\chi}t\right),
\end{equation}
showing Rabi oscillations between $|0\rangle$ and $|4\rangle$ with a frequency of
\begin{equation}
\omega_{04}=\frac{\sqrt{6}p^2}{2\chi}.
\end{equation}

\subsection{Perturbation calculation for the $|0\rangle, |6\rangle$ degeneracy}

At third order, we similarly obtain
\begin{equation}
   P_{0\rightarrow 6}(t) = |\langle 6|e^{-iHt}|0\rangle|^2 =\frac{1}{2} \left(1-\cos\frac{3\sqrt{5}p^3}{2\chi^2}t\right),
\end{equation}
demonstrating Rabi oscillations between $|0\rangle$ and $|6\rangle$ with frequency of
\begin{equation}
\omega_{06}=\frac{3\sqrt{5}p^3}{4\chi^2}.
\end{equation}


\ym{So} these calculations reveal that even when the direct transition matrix element $\langle n|V|0\rangle$ vanishes (e.g., for $n=4,6$), higher-order perturbative processes enable the occurrence of Rabi oscillations. 
\tana{As $n=2,4,6,…$ increases, the corresponding order of the perturbation scales as $p$ for first order, $p^2/\chi$ for second order, and $p^3/\chi^2$ for third order. Consequently, the resulting frequency is proportional to $p^{n/2}/\chi^{(n/2)-1}$.}

\section{Numerical verification of Rabi oscillations at higher-order degeneracies}
\onecolumngrid

\vspace{15pt}
\begin{figure}[htbp]
    \centering

    \begin{subfigure}[t]{0.32\textwidth}
        \centering
        \includegraphics[width=\linewidth]{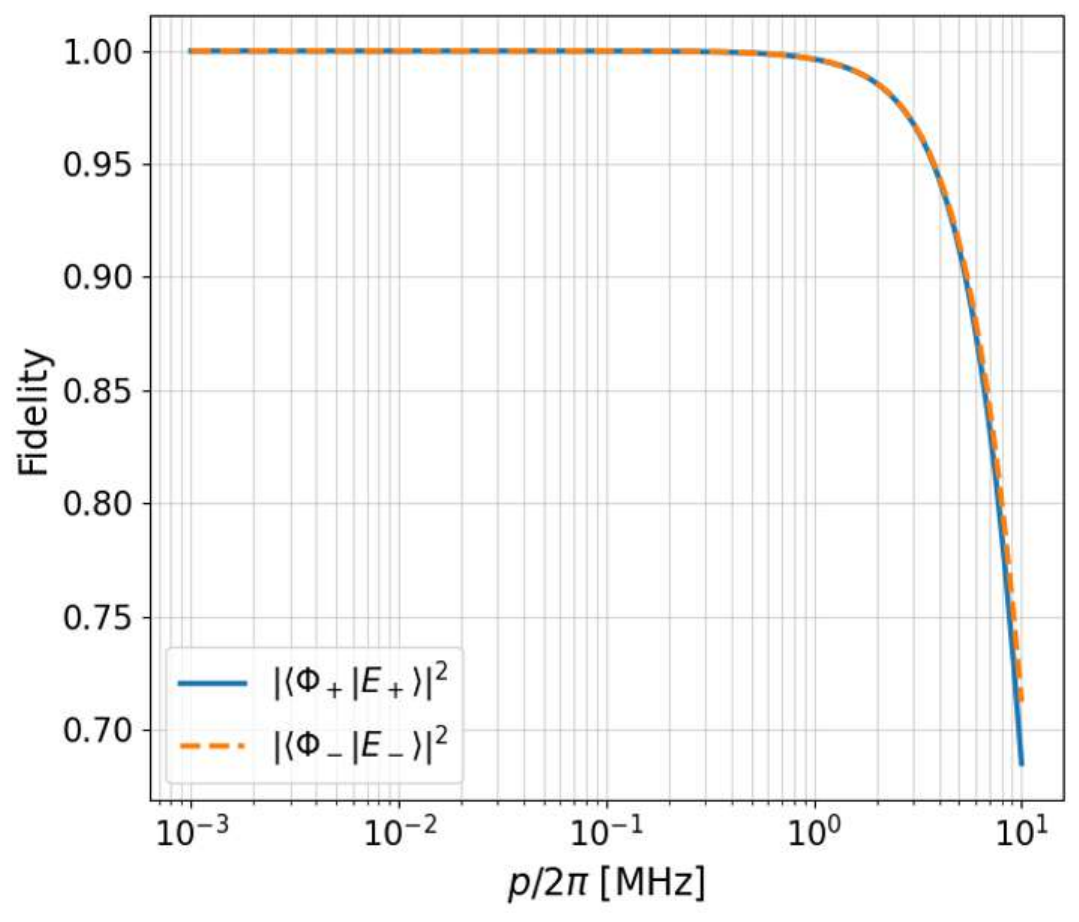}
        \caption{
        Fidelity as a function of the parametric drive amplitude for the $|0\rangle$--$|8\rangle$ degeneracy.
        }
        \label{08p_eig}
    \end{subfigure}
    \hfill
    \begin{subfigure}[t]{0.32\textwidth}
        \centering
        \includegraphics[width=\linewidth]{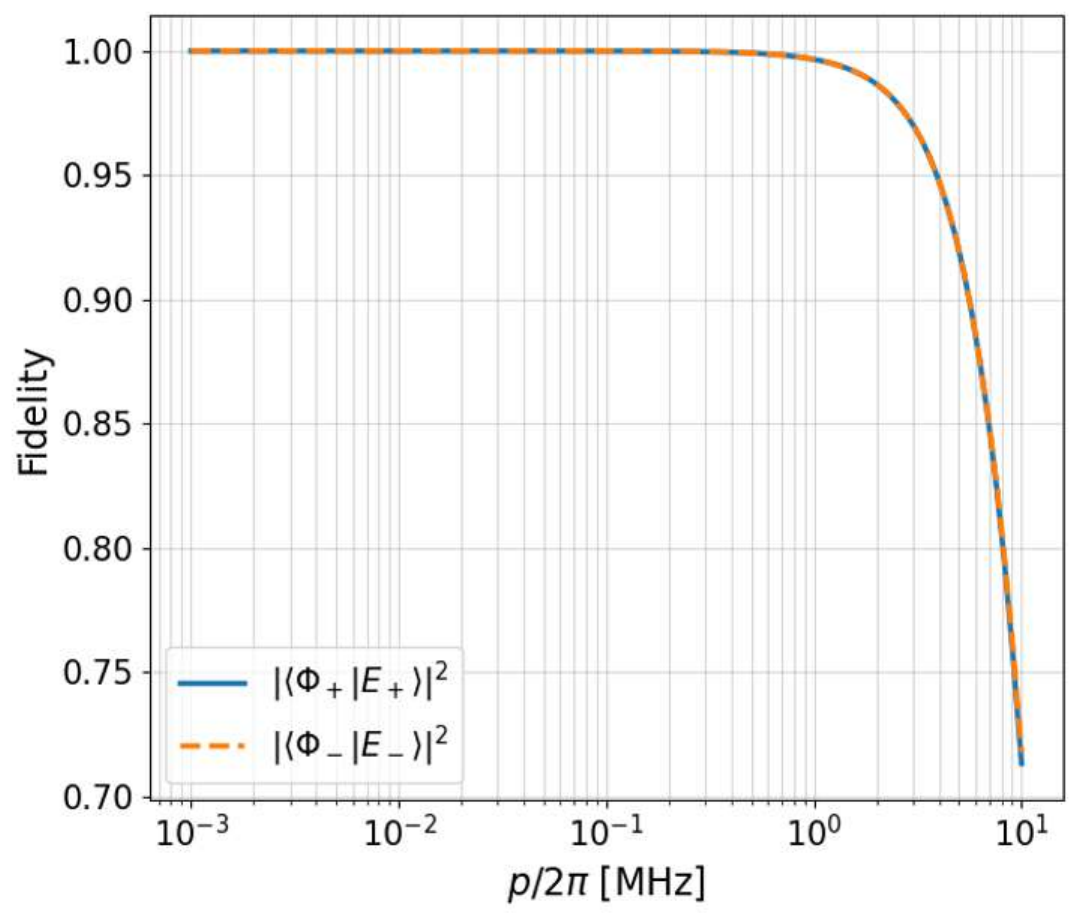}
        \caption{
        Fidelity as a function of the parametric drive amplitude for the $|0\rangle$--$|10\rangle$ degeneracy.
         }
        \label{010p_eig}
    \end{subfigure}
    \hfill
    \begin{subfigure}[t]{0.32\textwidth}
        \centering
        \includegraphics[width=\linewidth]{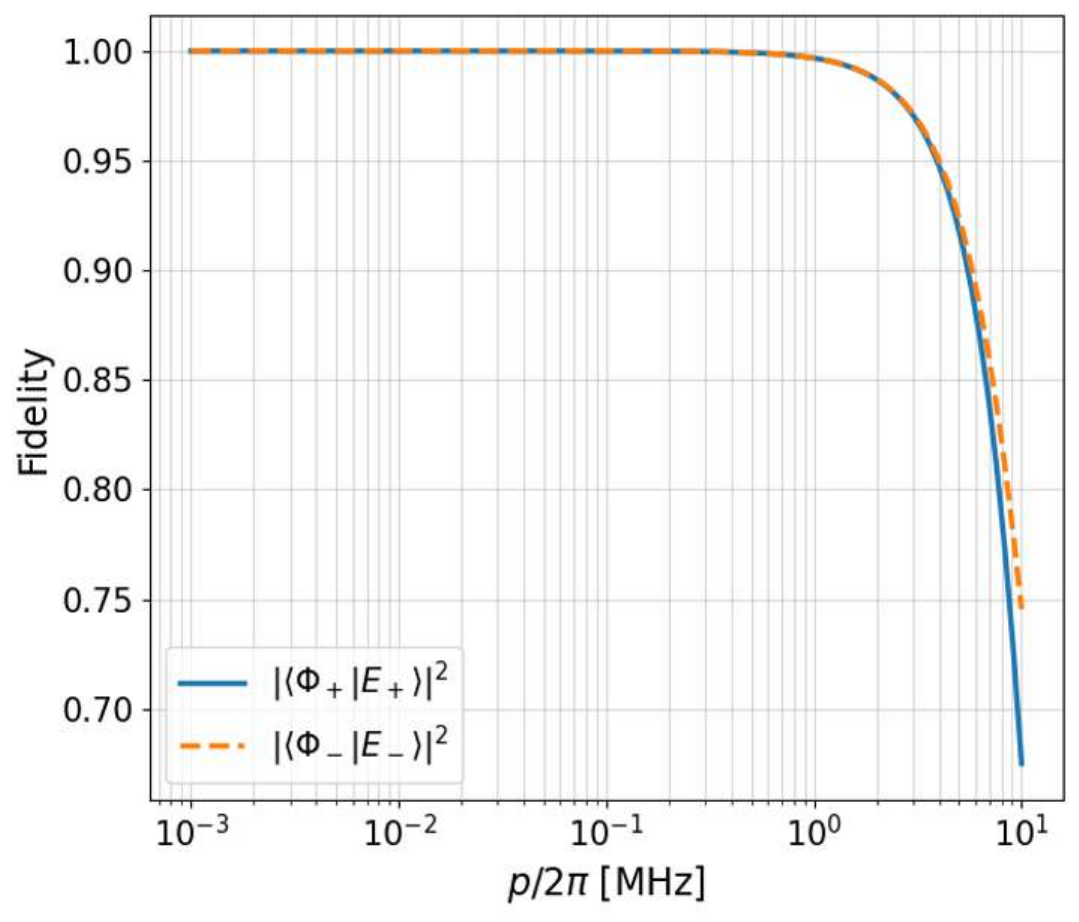}
        \caption{
        Fidelity as a function of the parametric drive amplitude for the $|0\rangle$--$|12\rangle$ degeneracy.
         }
        \label{012p_eig}
    \end{subfigure}

    \caption{
    Numerical calculation of the eigenstates $|E_+\rangle$ and $|E_-\rangle$ of the Hamiltonian in Eq.~\eqref{h_kpo} for the $|0\rangle$--$|n\rangle$ degeneracy.  
    The fidelities $|\langle \phi_+ | E_+ \rangle|^2$ and $|\langle \phi_- | E_- \rangle|^2$ are plotted as a function of the parametric drive amplitude $p/2\pi$ [MHz].  
    The nonlinearity is set to $\chi/2\pi = 18$~MHz.
    }
    \label{figall04}
\end{figure}

\begin{figure}[htbp]
    \centering

    \begin{subfigure}[t]{0.32\textwidth}
        \centering
        \includegraphics[width=\linewidth]{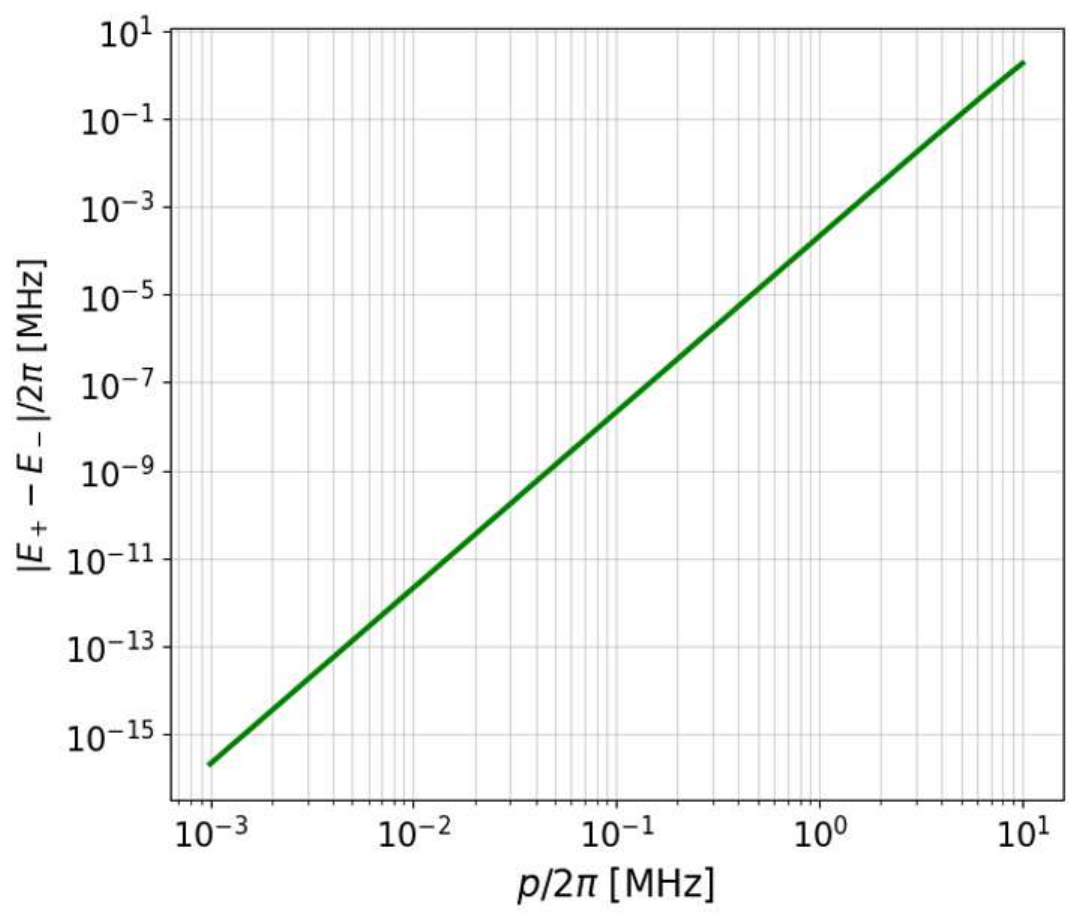}
        \caption{
        Energy splitting as a function of the parametric drive amplitude for the $|0\rangle$--$|8\rangle$ degeneracy.
        }
        \label{08ene}
    \end{subfigure}
    \hfill
    \begin{subfigure}[t]{0.32\textwidth}
        \centering
        \includegraphics[width=\linewidth]{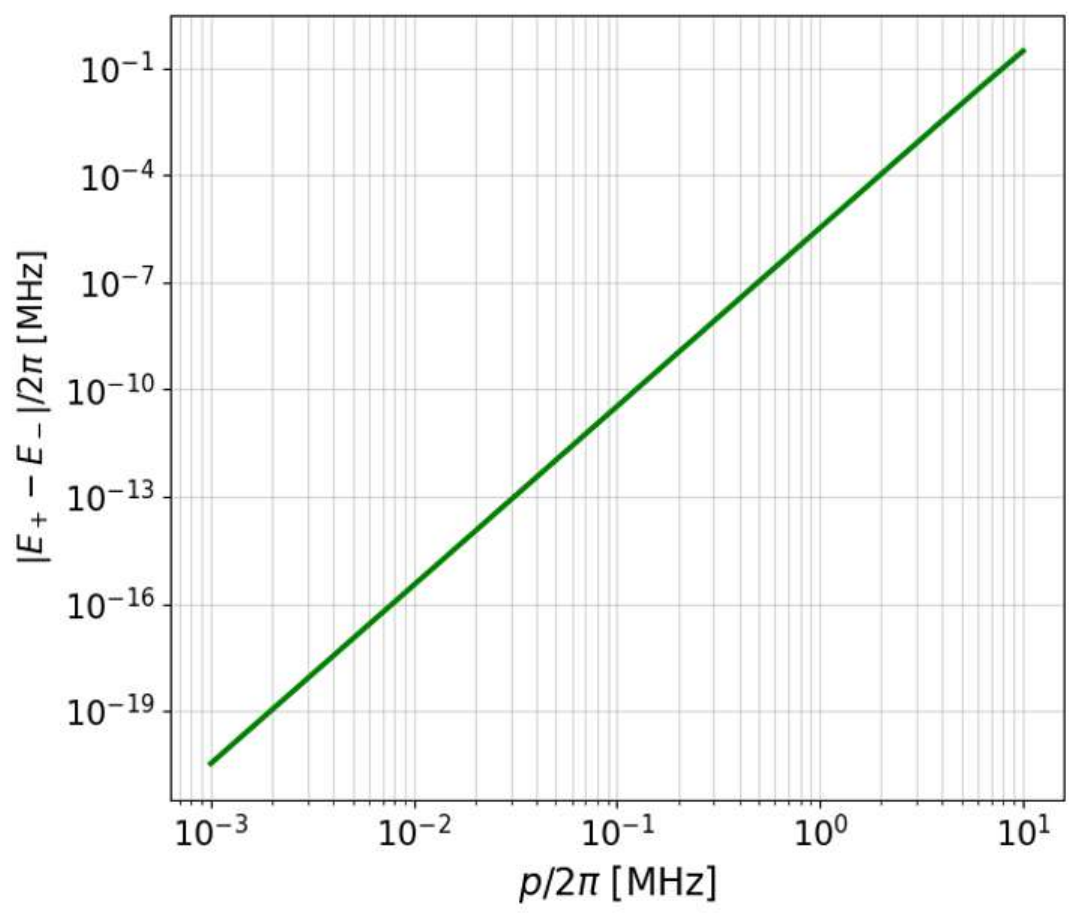}
        \caption{
        Energy splitting as a function of the parametric drive amplitude for the $|0\rangle$--$|10\rangle$ degeneracy.
         }
        \label{010ene}
    \end{subfigure}
    \hfill
    \begin{subfigure}[t]{0.32\textwidth}
        \centering
        \includegraphics[width=\linewidth]{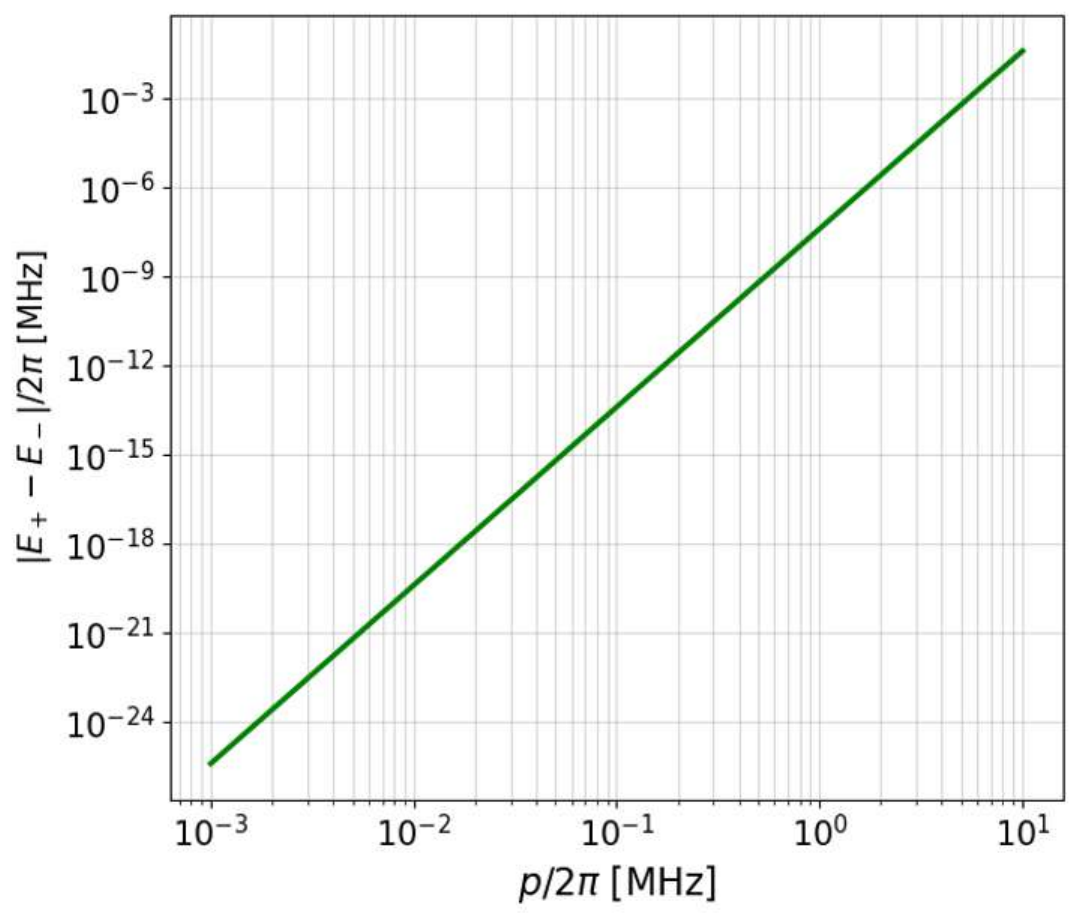}
        \caption{
        Energy splitting as a function of the parametric drive amplitude for the $|0\rangle$--$|12\rangle$ degeneracy.
         }
        \label{012ene}
    \end{subfigure}

    \caption{
    Numerical calculation of the energy levels $E_+$ and $E_-$ of the Hamiltonian in Eq.~\eqref{h_kpo} for the $|0\rangle$--$|n\rangle$ degeneracy.  
    The energy difference $(E_+ - E_-)/2\pi$ [MHz] is plotted as a function of the parametric drive amplitude $p/2\pi$ [MHz].  
    The nonlinearity is set to $\chi/2\pi = 18$~MHz.
    }
    \label{ene_dif}
\end{figure}

\begin{figure}[htbp]
    \centering

    \begin{subfigure}[t]{0.32\textwidth}
        \centering
        \includegraphics[width=\linewidth]{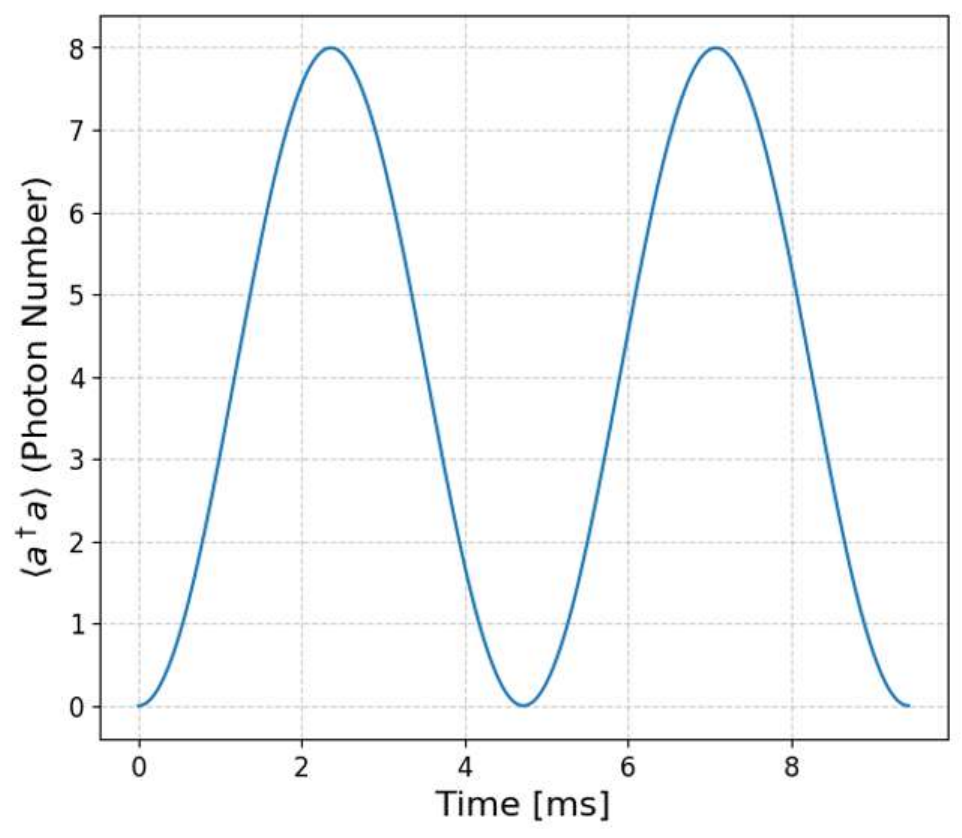}
        \caption{
        Time evolution of the photon number expectation value for the $|0\rangle$--$|8\rangle$ degeneracy.
        }
        \label{rch08}
    \end{subfigure}
    \hfill
    \begin{subfigure}[t]{0.32\textwidth}
        \centering
        \includegraphics[width=\linewidth]{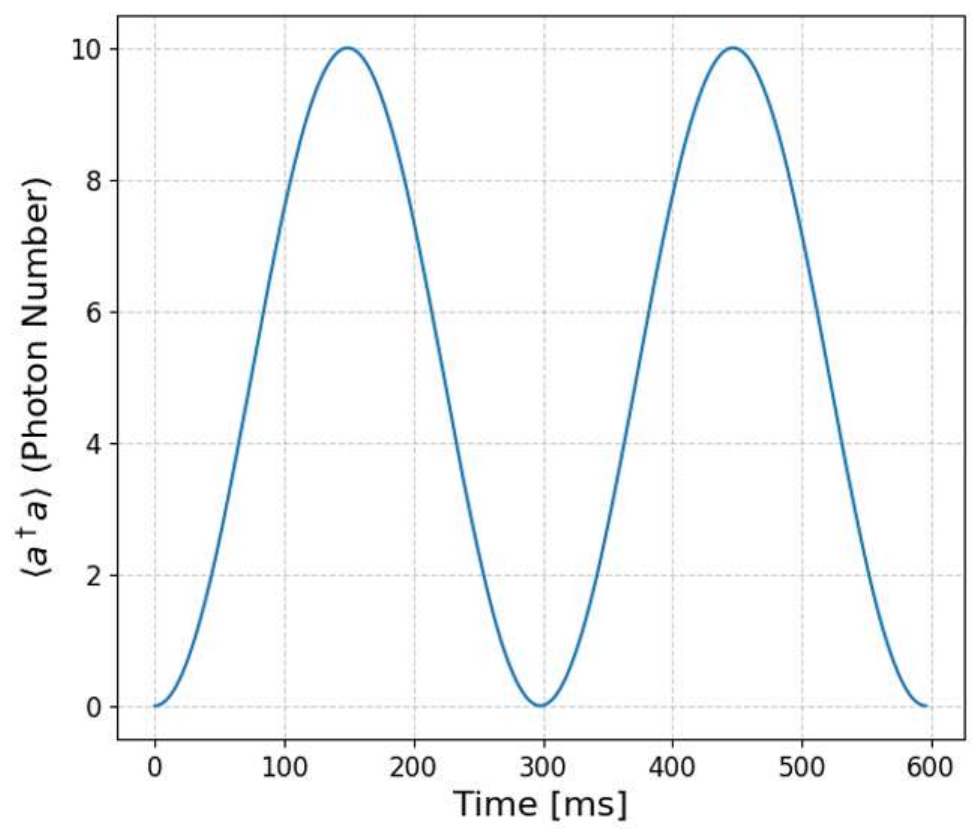}
        \caption{
        Time evolution of the photon number expectation value for the $|0\rangle$--$|10\rangle$ degeneracy.
         }
        \label{rch010}
    \end{subfigure}
    \hfill
    \begin{subfigure}[t]{0.32\textwidth}
        \centering
        \includegraphics[width=\linewidth]{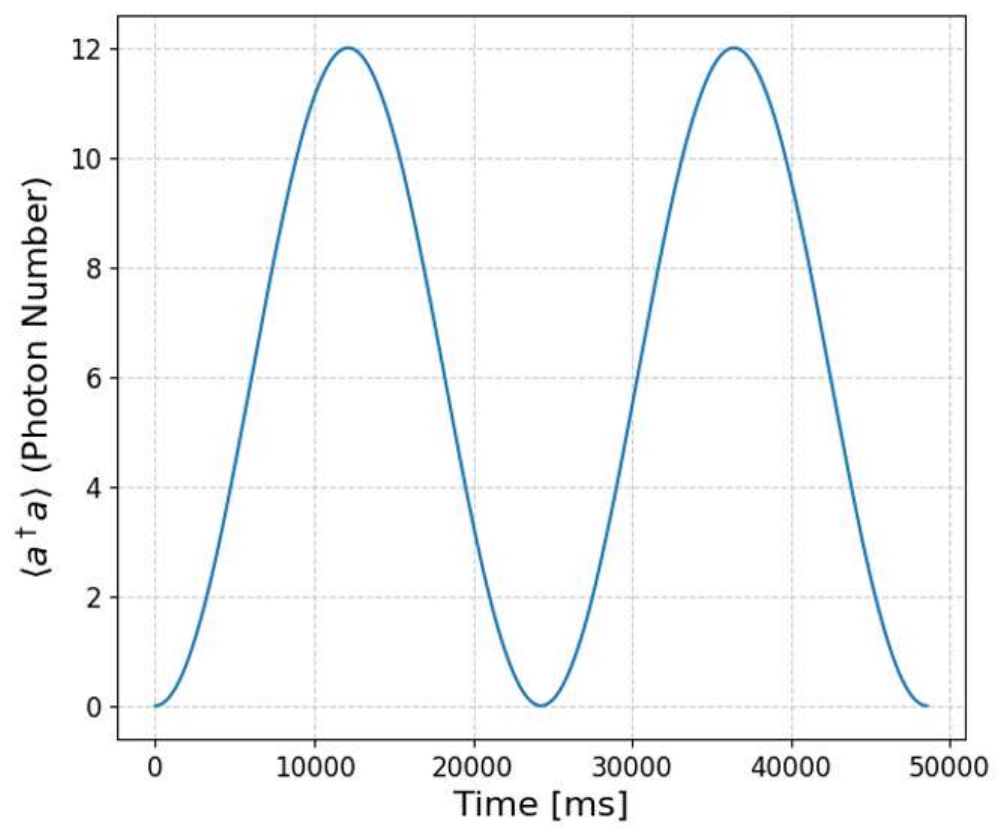}
        \caption{
        Time evolution of the photon number expectation value for the $|0\rangle$--$|12\rangle$ degeneracy.
         }
        \label{rch012}
    \end{subfigure}

    \caption{
    Time evolution of the photon number expectation value $\langle a^{\dagger}a\rangle$ in the KPO system for the $|0\rangle$--$|n\rangle$ degeneracy, showing the presence of Rabi oscillations.  
    The horizontal axis represents time $t$ [µs], and the vertical axis represents $\langle a^{\dagger}a\rangle$.  
    The parameters are set to $\chi/2\pi = 18$~MHz and $p/2\pi = 1$~MHz.
    }
    \label{rch}
\end{figure}
\twocolumngrid
 
In this 
\ym{section}, we numerically verify that Rabi oscillations also occur for higher-order cases ($N = 8, 10, 12$).  
\ym{While we analytically showed that Rabi oscillations occur at the degeneracies between $|0\rangle$ and $|2\rangle$, $|0\rangle$ and $|4\rangle$, and $|0\rangle$ and $|6\rangle$ in the previous section, it is difficult to perform analytical calculations for the higher-order cases, and so we use numerical approaches. }

We consider the case where a degeneracy occurs between $|0\rangle$ and $|n\rangle$ \ym{for $n=8,10,12$}.  
We define the superposition states composed of the vacuum and the $n$-photon Fock state as
\begin{equation}
|\phi_+\rangle = \frac{|0\rangle + |n\rangle}{\sqrt{2}}, \quad
|\phi_-\rangle = \frac{|0\rangle - |n\rangle}{\sqrt{2}}.
\end{equation}
The Hamiltonian given by Eq.~\eqref{h_kpo} is numerically diagonalized to obtain its eigenvectors.  
Among them, we identify the eigenvectors $|E_+\rangle$ and $|E_-\rangle$ that have the highest fidelity with $|\phi_+\rangle$ and $|\phi_-\rangle$, respectively.  
We then plot the fidelity values $|\langle \phi_+ | E_+ \rangle|^2$ and $|\langle \phi_- | E_- \rangle|^2$ as a function of the drive amplitude $p/2\pi$ [MHz] (see Figs.~\ref{08p_eig}, \ref{010p_eig}, and \ref{012p_eig}).  
For $n = 8, 10, 12$, it is numerically confirmed that $|E_+\rangle$ ($|E_-\rangle$) coincides with $|\phi_+\rangle$ ($|\phi_-\rangle$) \ym{for weak parametric driving}

Furthermore, to investigate the Rabi frequency, we plot the energy difference $(E_+ - E_-)/2\pi$ [MHz] between the eigenenergies $E_+$ and $E_-$ as a function of the drive amplitude $p/2\pi$ [MHz] (Figs.~\ref{08ene}, \ref{010ene}, and \ref{012ene}).  
This energy splitting corresponds to the observed Rabi frequency.  

In addition, by fixing the drive amplitude at $p/2\pi = 1$~MHz, we solve the Schrödinger equation numerically.  
We plot the photon number expectation value $\langle a^{\dagger}a\rangle$ as a function of time for $n = 8, 10, 12$, and confirm that Rabi oscillations indeed occur (Figs.~\ref{rch08}, \ref{rch010}, and \ref{rch012}).

\section{Observation of Rabi Oscillations via Numerical Simulation}
\onecolumngrid

\begin{figure}[htbp]
    \centering

    \begin{subfigure}[t]{0.32\textwidth}
        \centering
        \includegraphics[width=\linewidth]{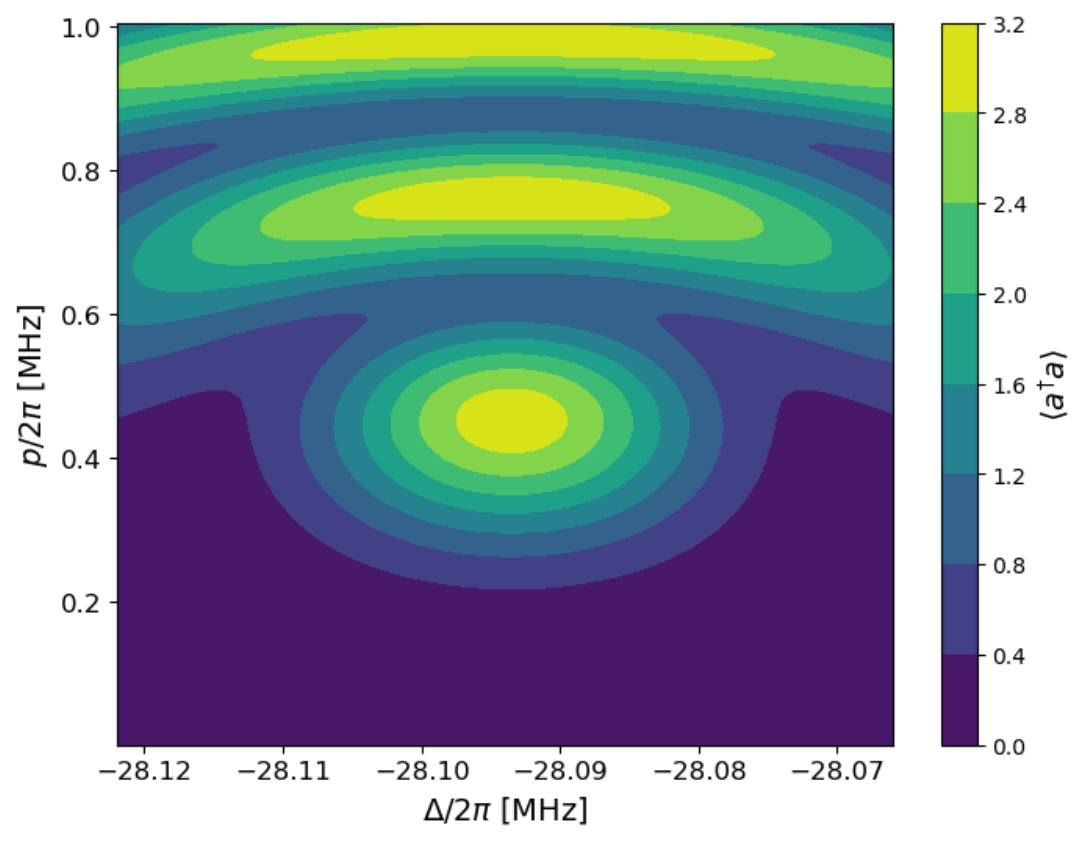}
        \caption{
        Plot at $t=10\,\mu$s.
        }
        \label{04t005}
    \end{subfigure}
    \hfill
    \begin{subfigure}[t]{0.32\textwidth}
        \centering
        \includegraphics[width=\linewidth]{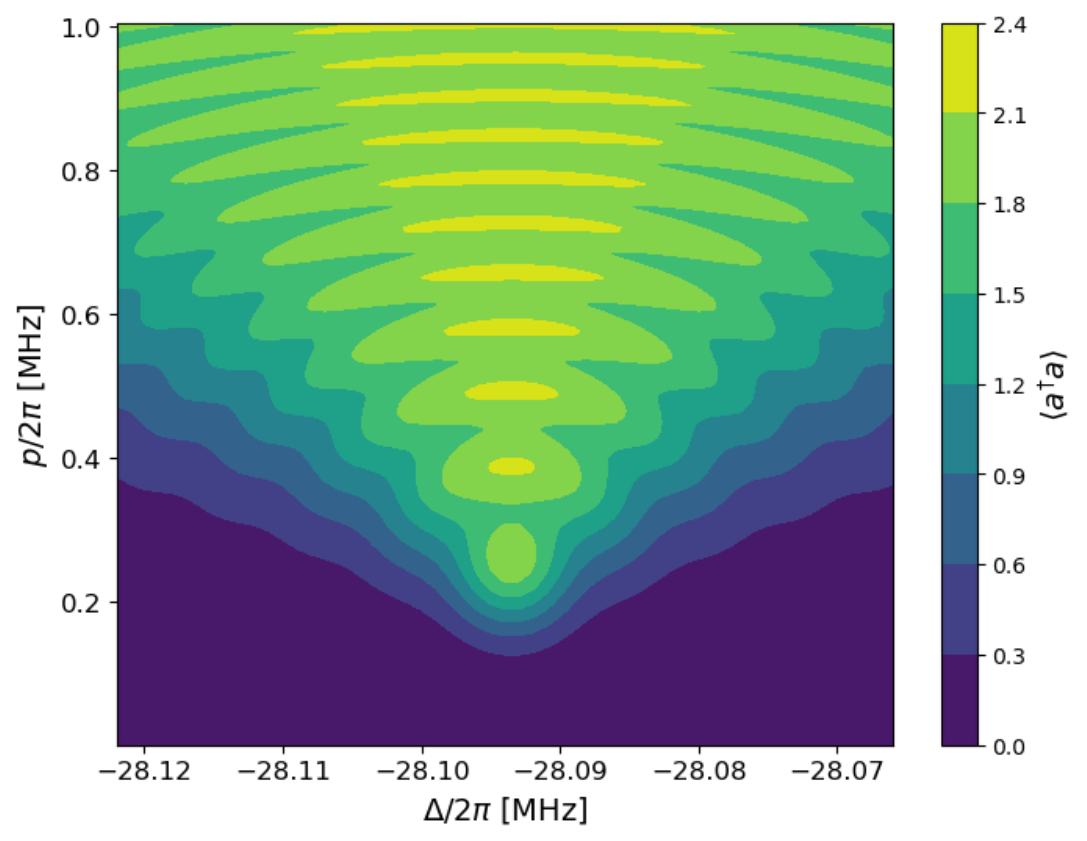}
        \caption{
        Plot at $t=40\,\mu$s.
         }
        \label{04t02}
    \end{subfigure}
    \hfill
    \begin{subfigure}[t]{0.32\textwidth}
        \centering
        \includegraphics[width=\linewidth]{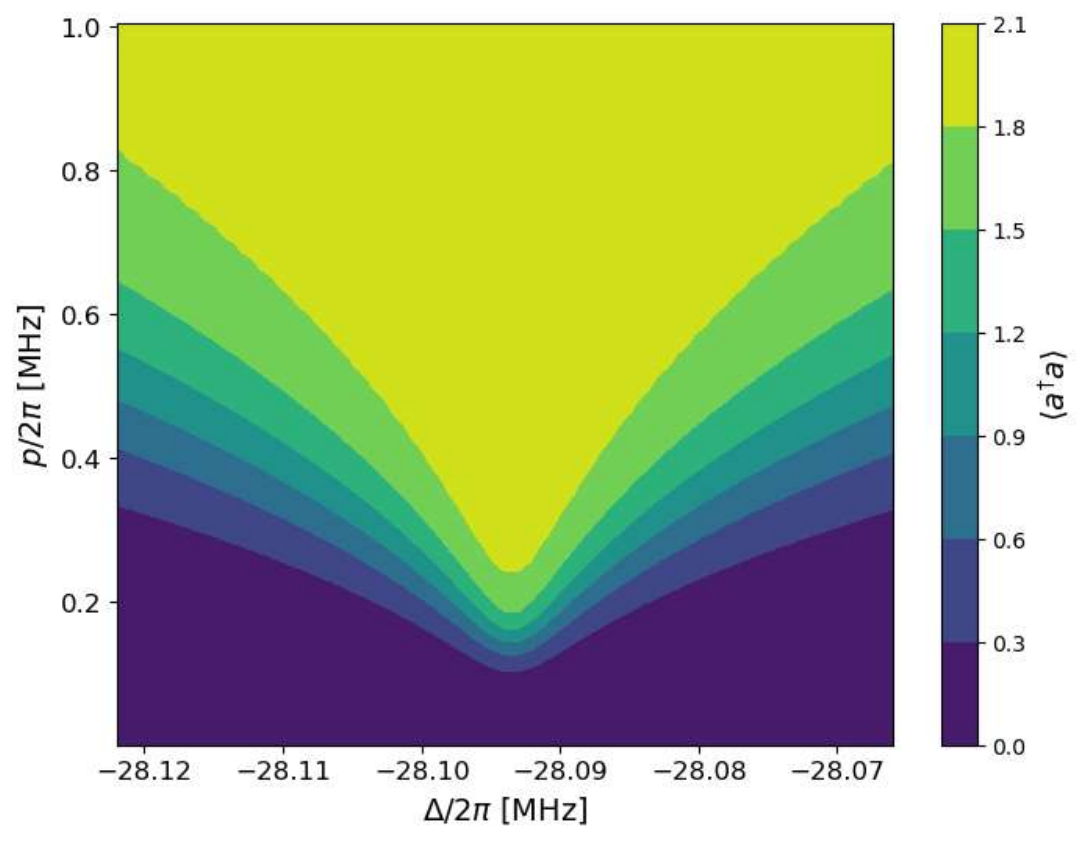}
        \caption{
        Plot at $t=100\,\mu$s.
        }
        \label{04t100}
    \end{subfigure}
    
    \caption{
    The master equation was solved, and the expectation value of the photon number of the KPO was plotted near the degeneracy condition between $|0\rangle$ and $|4\rangle$. 
    The $x$-axis represents the detuning $\Delta/2\pi$ [MHz], the $y$-axis represents the parametric drive $p/2\pi$ [MHz], and the $z$-axis shows the photon number expectation value $\langle a^{\dagger}a\rangle$. 
    The parameters are set to the dissipation rate $\kappa/2\pi=5\times10^{-3}$ MHz and the nonlinearity $\chi/2\pi=18.729$ MHz.
 }
    \label{04_3dim}
\end{figure}
\begin{figure}[htbp]
    \centering

    \begin{subfigure}[t]{0.25\textwidth}
        \centering
        \includegraphics[width=\linewidth]{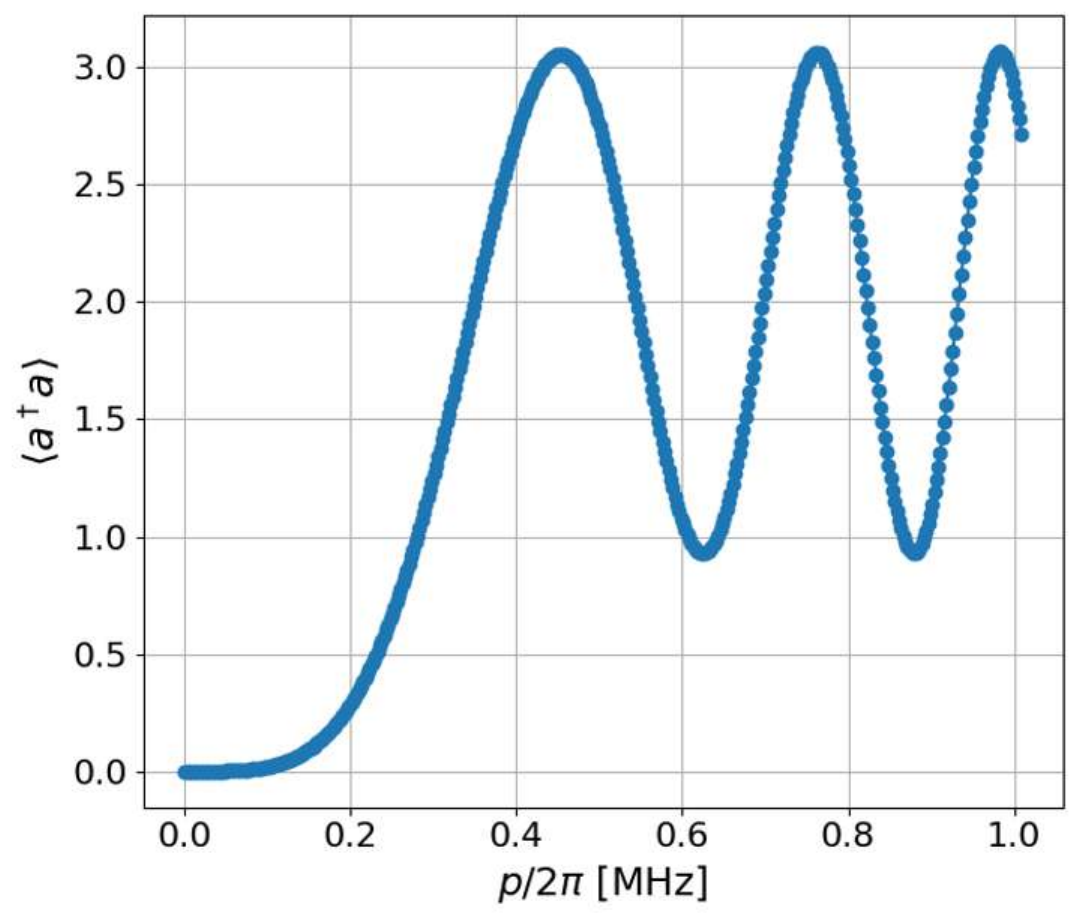}
        \caption{
        Plot at $t=10\,\mu$s. Rabi oscillations are observed.
        }
        \label{04t005r}
    \end{subfigure}
    \hfill
    \begin{subfigure}[t]{0.25\textwidth}
        \centering
        \includegraphics[width=\linewidth]{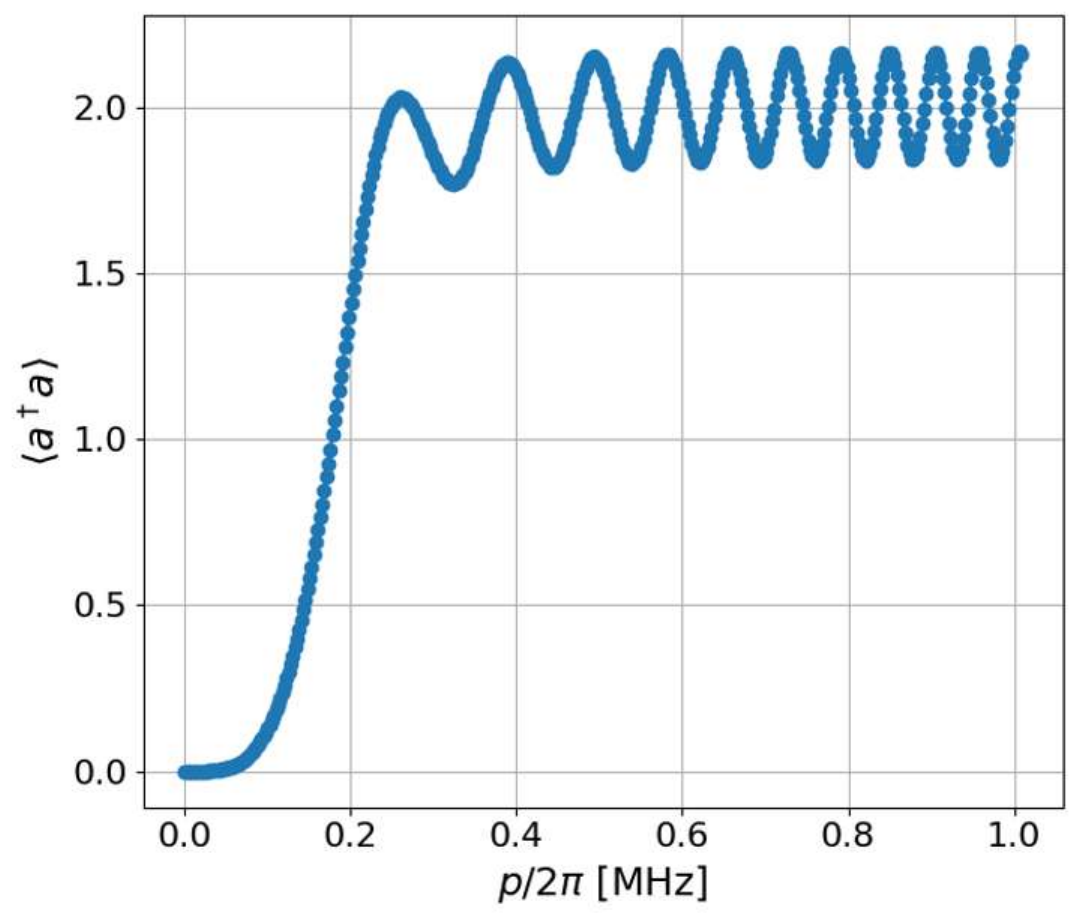}
        \caption{
        Plot at $t=40\,\mu$s. The effect of decoherence becomes stronger, and the oscillation amplitude weakens.
        }
        \label{04t20r}
    \end{subfigure}
    \hfill
    \begin{subfigure}[t]{0.25\textwidth}
        \centering
        \includegraphics[width=\linewidth]{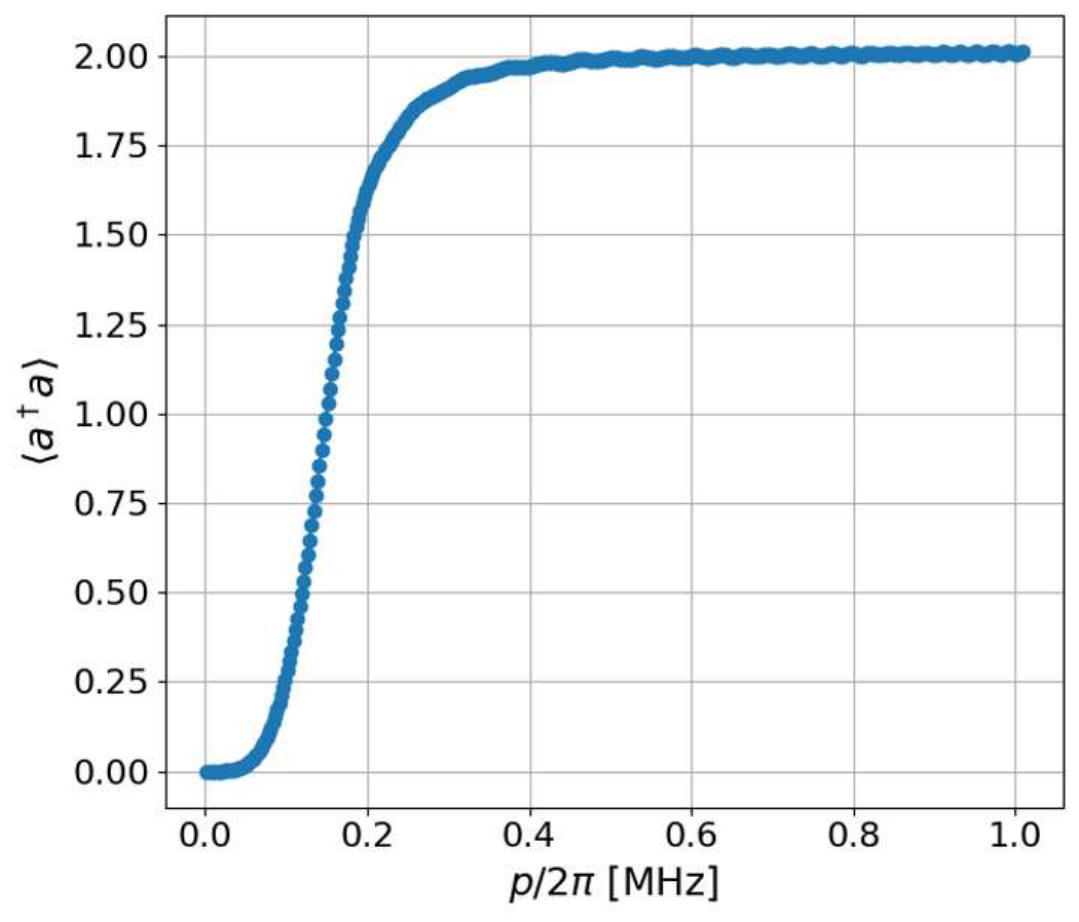}
        \caption{
        Plot at $t=100\,\mu$s. The system approaches a steady state.
        }
        \label{rabi04100}
    \end{subfigure}
    
    \caption{
    Time evolution of Rabi oscillations under the degeneracy condition between $|0\rangle$ and $|4\rangle$ at $\Delta/2\pi=-28.0935$ MHz. 
    The $x$-axis represents the parametric drive $p/2\pi$ [MHz], and the $z$-axis represents the photon number expectation value $\langle a^{\dagger}a\rangle$. 
    The parameters are set to the dissipation rate $\kappa/2\pi=5\times10^{-3}$ MHz and the nonlinearity $\chi/2\pi=18.729$ MHz.
 }
    \label{rabi04}
\end{figure}
\begin{figure}[htbp]
    \centering

    \begin{subfigure}[t]{0.32\textwidth}
        \centering
        \includegraphics[width=\linewidth]{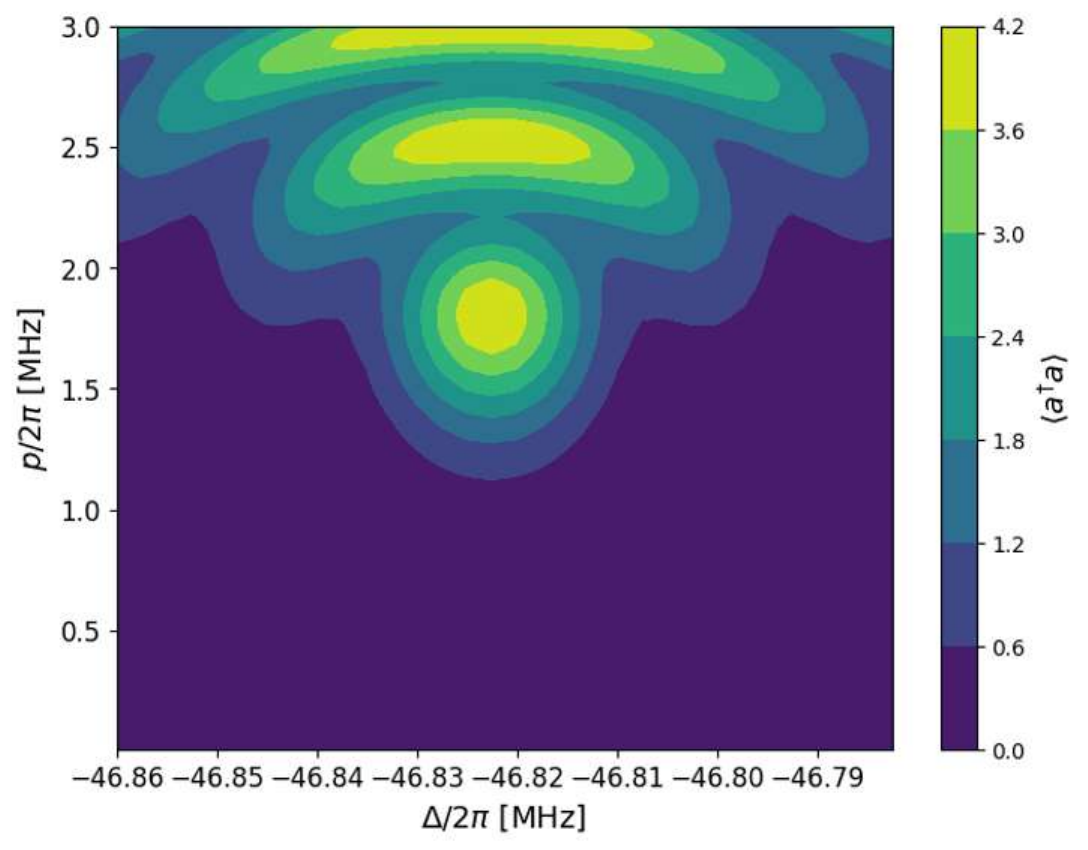}
        \caption{
        Plot at $t=10\,\mu$s.
        }
        \label{fig:deg05}
    \end{subfigure}
    \hfill
    \begin{subfigure}[t]{0.32\textwidth}
        \centering
        \includegraphics[width=\linewidth]{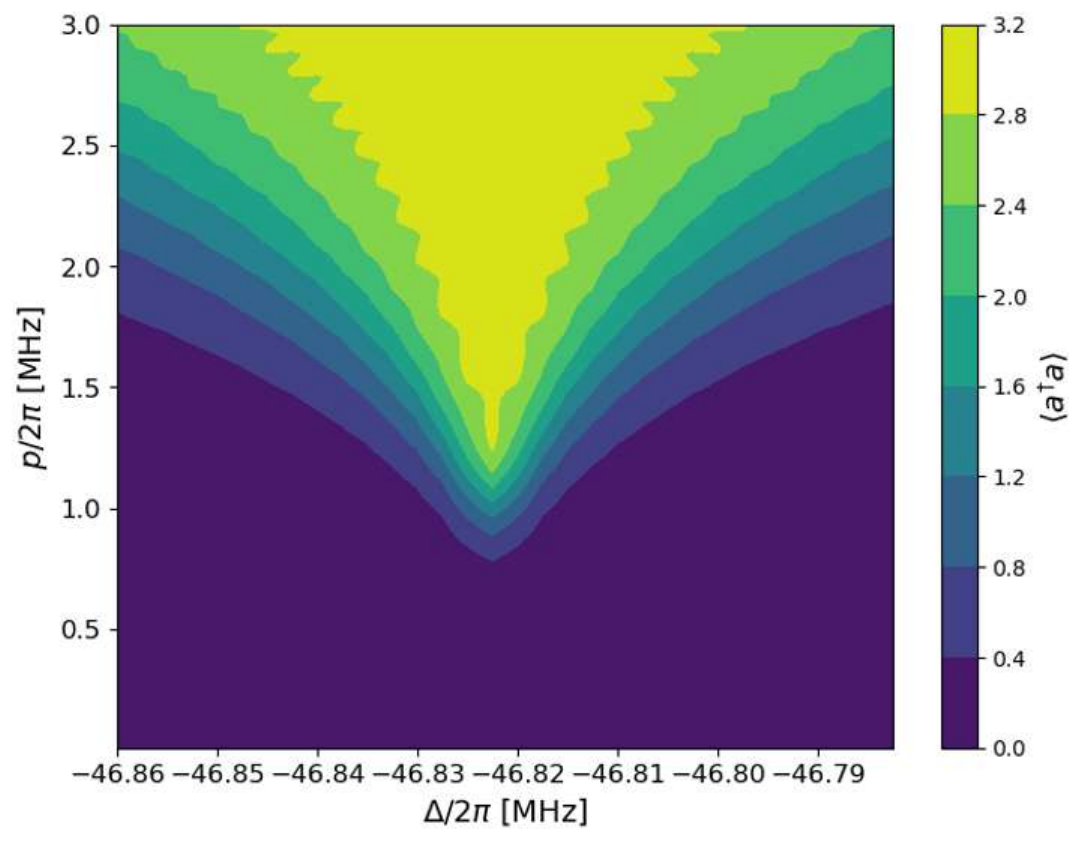}
        \caption{
        Plot at $t=40\,\mu$s.
        }
        \label{fig06_40}
    \end{subfigure}
    \hfill
    \begin{subfigure}[t]{0.32\textwidth}
        \centering
        \includegraphics[width=\linewidth]{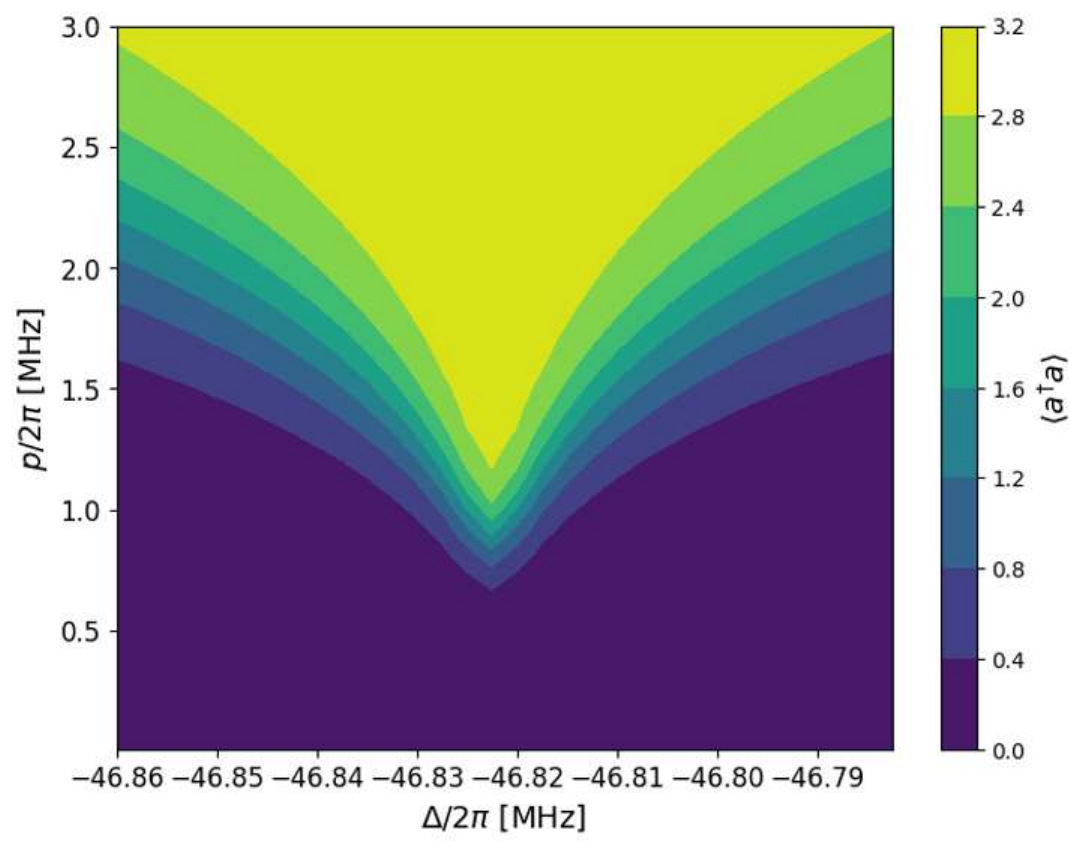}
        \caption{
        Plot at $t=100\,\mu$s.
        }
        \label{fig06_100}
    \end{subfigure}

    \caption{
    The master equation was solved, and the photon number expectation value of the KPO was plotted near the degeneracy condition between $|0\rangle$ and $|6\rangle$. 
    The $x$-axis represents the detuning $\Delta/2\pi$ [MHz], the $y$-axis represents the normalized parametric drive $p/\chi$, and the $z$-axis represents the photon number expectation value $\langle a^{\dagger}a\rangle$. 
    The parameters are set to the dissipation rate $\kappa/2\pi=5\times10^{-3}$ MHz and the nonlinearity $\chi/2\pi=18.729$ MHz.
 }
    \label{06_3dim}
\end{figure}
\begin{figure}[htbp]
    \centering

    \begin{subfigure}[t]{0.25\textwidth}
        \centering
        \includegraphics[width=\linewidth]{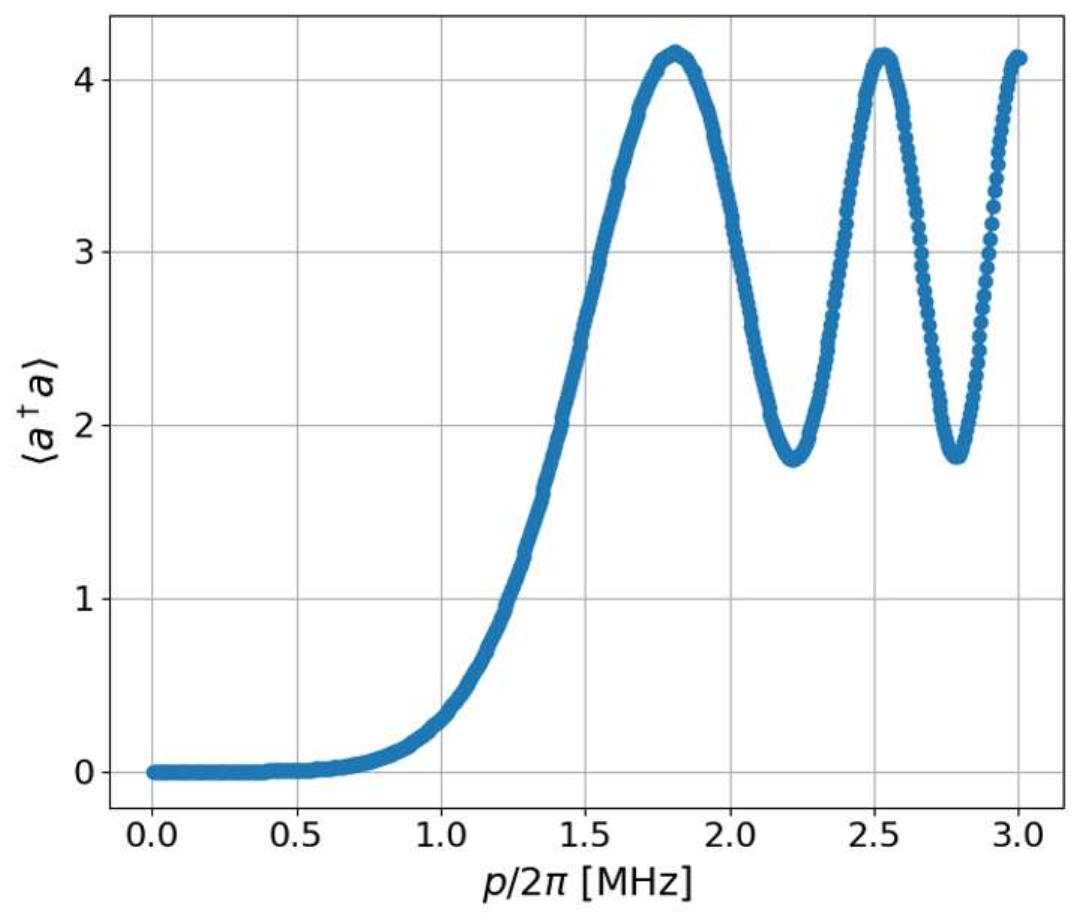}
        \caption{
        Plot at $t=10\,\mu$s. Rabi oscillations can be observed.
        }
        \label{rabi06005}
    \end{subfigure}
    \hfill
    \begin{subfigure}[t]{0.25\textwidth}
        \centering
        \includegraphics[width=\linewidth]{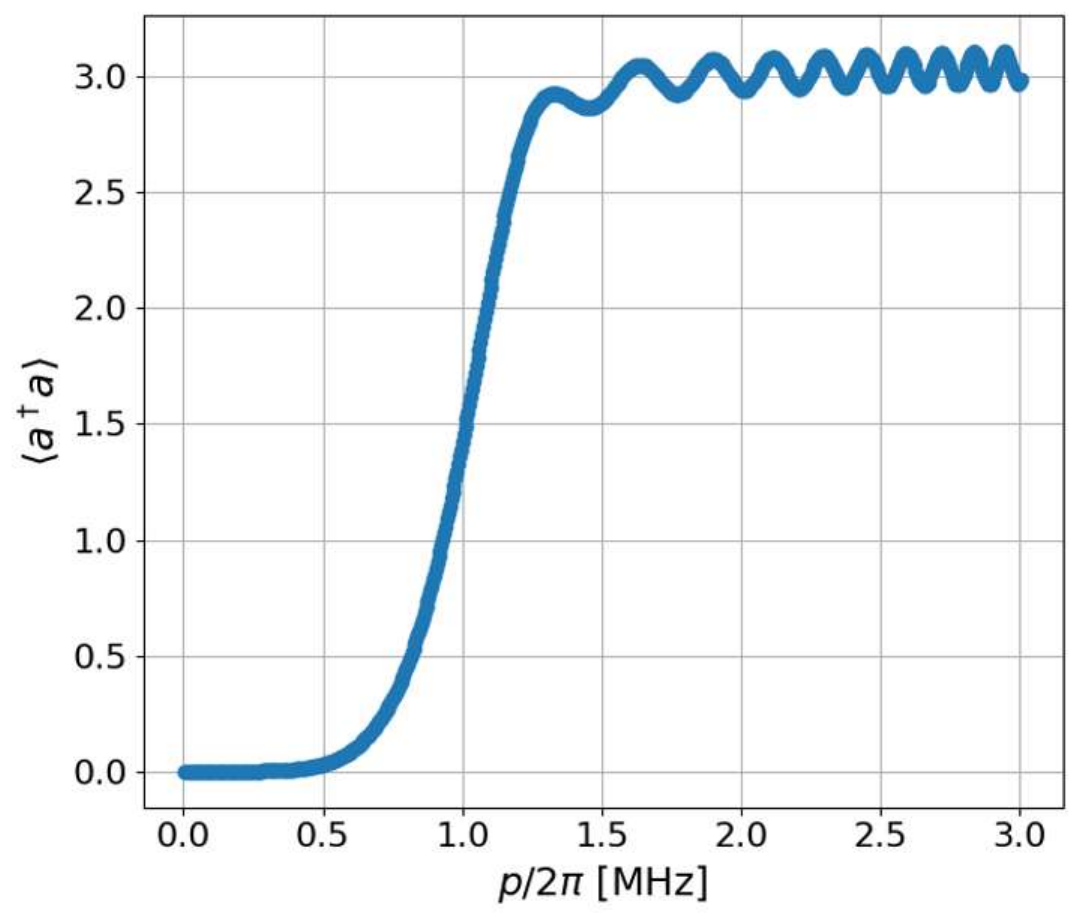}
        \caption{
        Plot at $t=40\,\mu$s. The effect of decoherence becomes stronger, and the oscillation amplitude weakens.
        }
        \label{rabi0602}
    \end{subfigure}
    \hfill
    \begin{subfigure}[t]{0.25\textwidth}
        \centering
        \includegraphics[width=\linewidth]{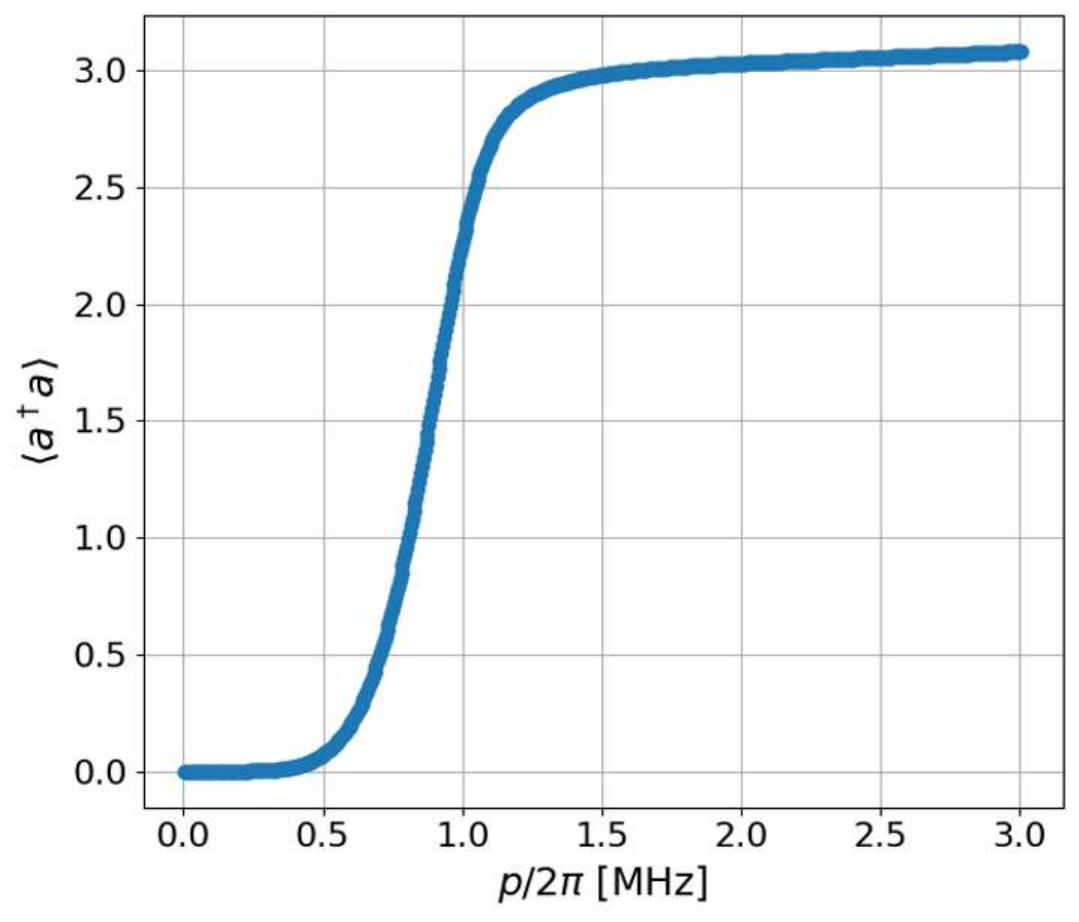}
        \caption{
        Plot at $t=100\,\mu$s. The system approaches a steady state.
        }
        \label{rabi061}
    \end{subfigure}

    \caption{
    Time evolution of Rabi oscillations under the degeneracy condition between $|0\rangle$ and $|6\rangle$ at $\Delta/2\pi=-46.8225$ MHz. 
    The $x$-axis represents the parametric drive $p/2\pi$ [MHz], and the $z$-axis represents the photon number expectation value $\langle a^{\dagger}a\rangle$. 
    The parameters are set to the dissipation rate $\kappa/2\pi=5\times10^{-3}$ MHz and the nonlinearity $\chi/2\pi=18.729$ MHz.
 }
    \label{rabi06}
\end{figure}
\begin{figure}[htbp]
    \centering

    \begin{subfigure}[t]{0.32\textwidth}
        \centering
        \includegraphics[width=\linewidth]{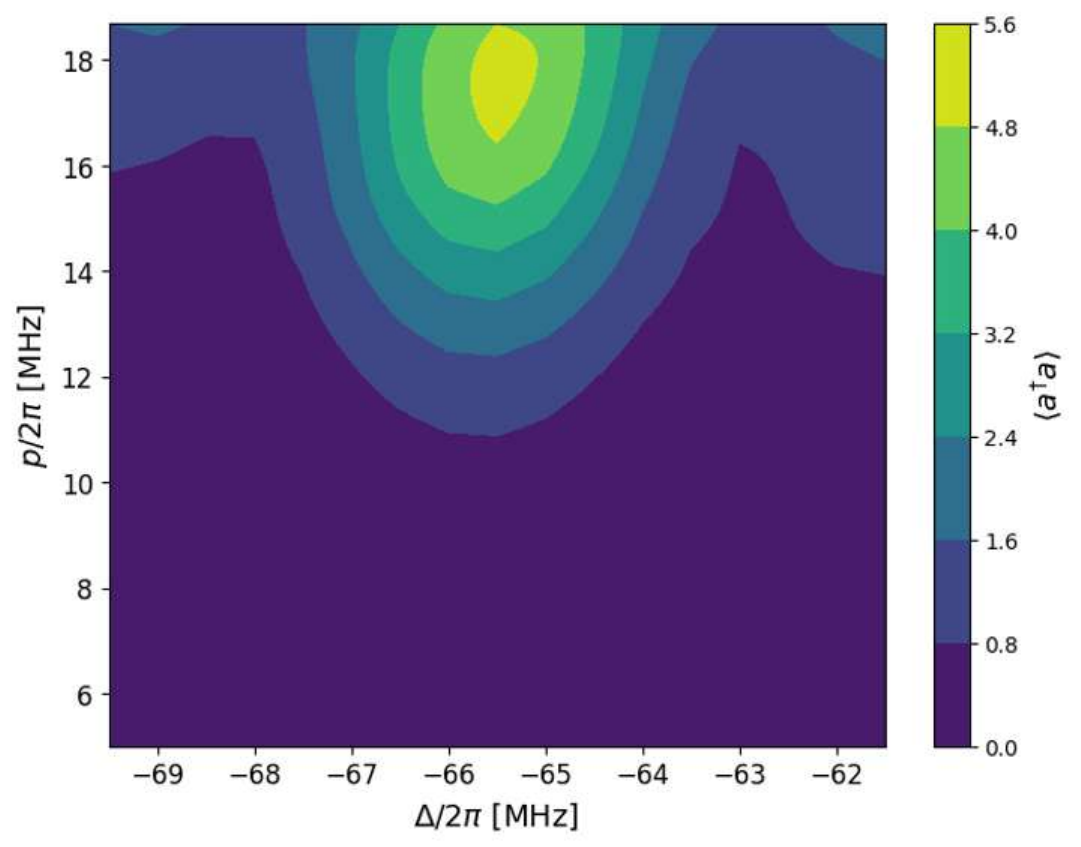}
        \caption{
        Plot at $t=0.05\,\mu$s.
        }
        \label{fig8_005}
    \end{subfigure}
    \hfill
    \begin{subfigure}[t]{0.32\textwidth}
        \centering
        \includegraphics[width=\linewidth]{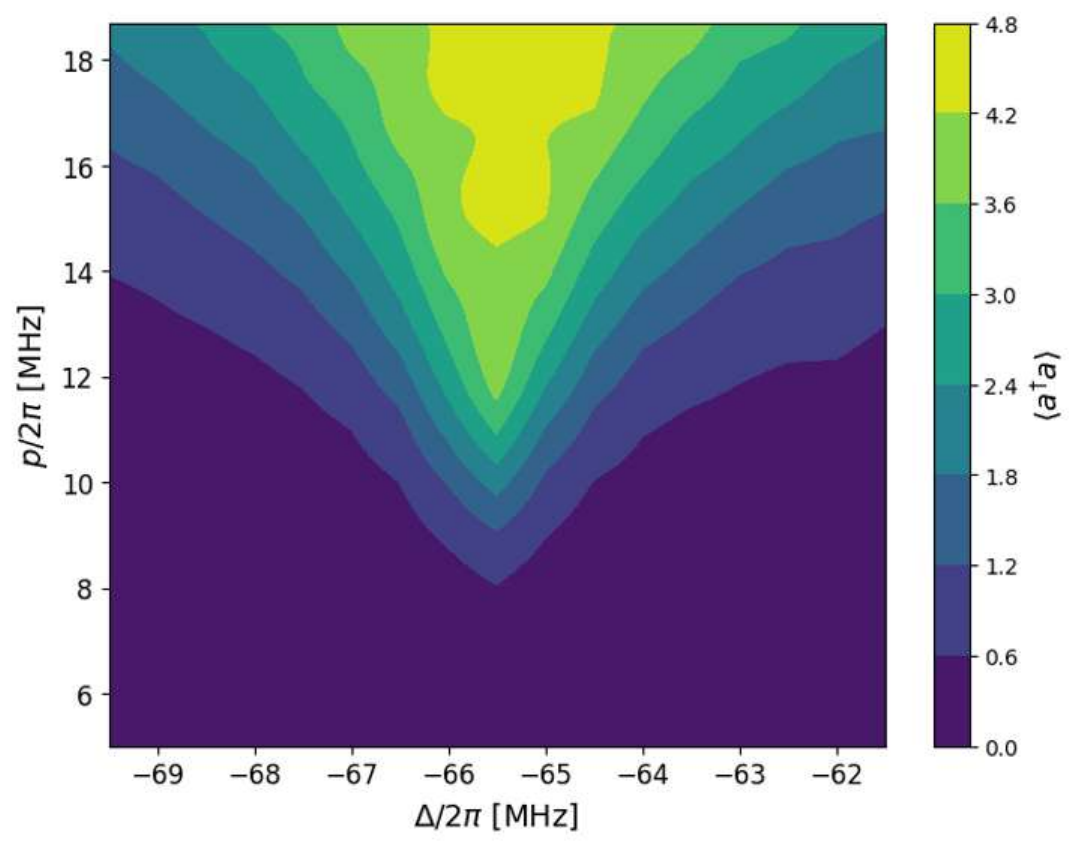}
        \caption{
        Plot at $t=0.2\,\mu$s.
         }
        \label{fig08_02}
    \end{subfigure}
    \hfill
    \begin{subfigure}[t]{0.32\textwidth}
        \centering
        \includegraphics[width=\linewidth]{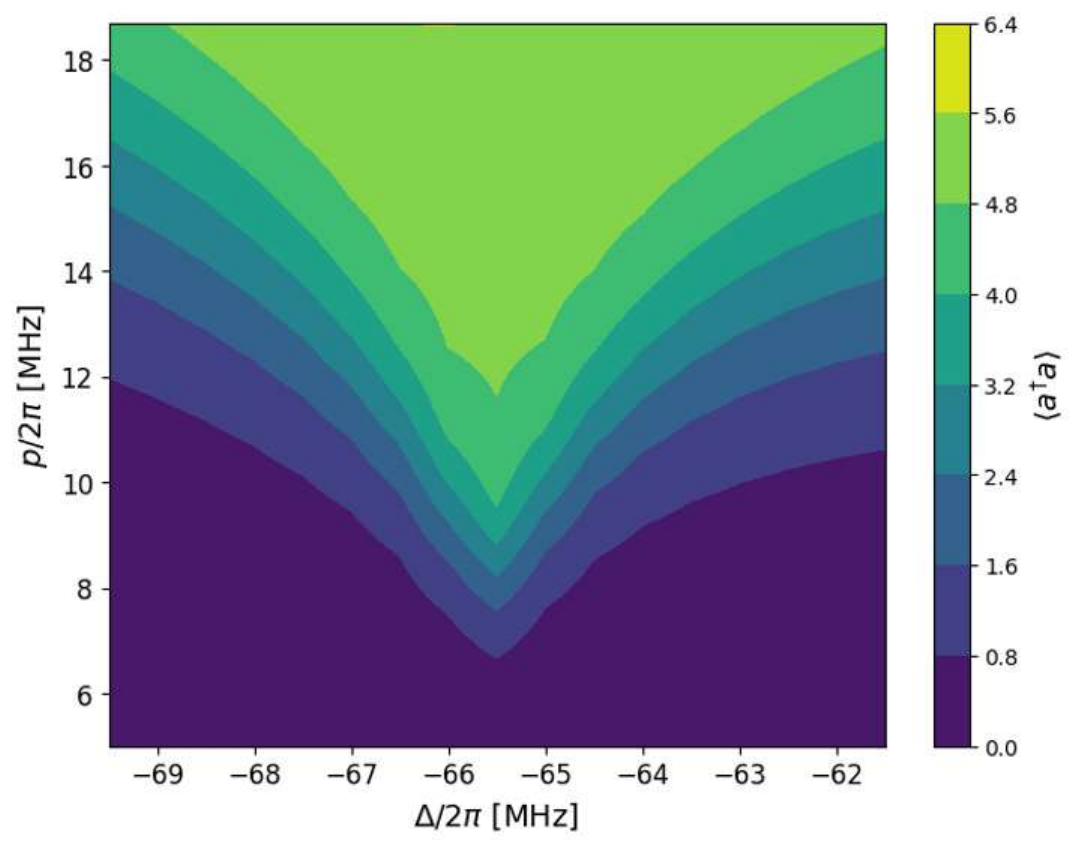}
        \caption{
        Plot at $t=1\,\mu$s.
         }
        \label{08t1}
    \end{subfigure}

    \caption{
    The master equation was solved, and the photon number expectation value of the KPO was plotted near the degeneracy condition between $|0\rangle$ and $|8\rangle$. 
    The $x$-axis represents the detuning $\Delta/2\pi$ [MHz], the $y$-axis represents the parametric drive $p/2\pi$ [MHz], and the $z$-axis represents the photon number expectation value $\langle a^{\dagger}a\rangle$. 
    The parameters are set to the dissipation rate $\kappa/2\pi=0.73$ MHz and the nonlinearity $\chi/2\pi=18.729$ MHz.
 }
    \label{08_3dim}
\end{figure}
\twocolumngrid

\clearpage
\onecolumngrid 

\begin{figure}[htbp]
    \centering

    \begin{subfigure}[t]{0.25\textwidth}
        \centering
        \includegraphics[width=\linewidth]{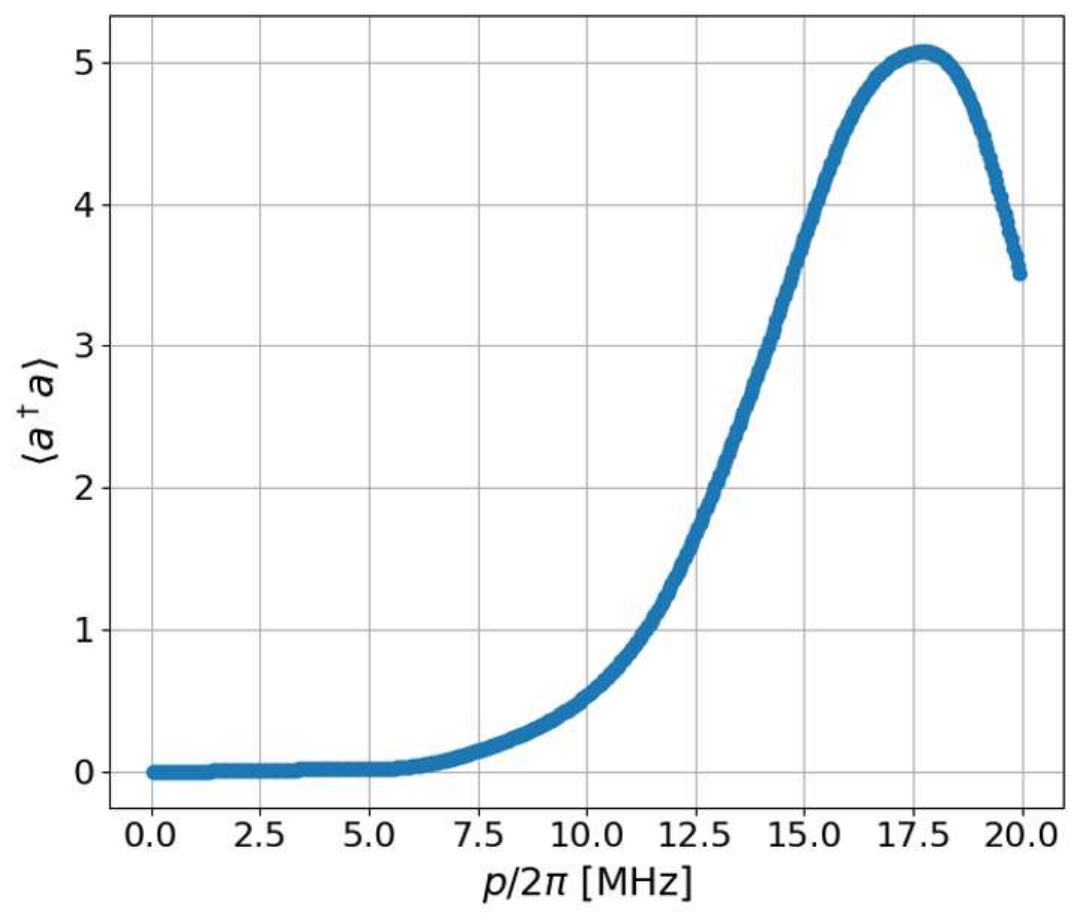}
        \caption{
        Plot at $t=0.05\,\mu$s. Rabi oscillations can be observed. 
           }
        \label{fig8_005_2d}
    \end{subfigure}
    \hfill
    \begin{subfigure}[t]{0.25\textwidth}
        \centering
        \includegraphics[width=\linewidth]{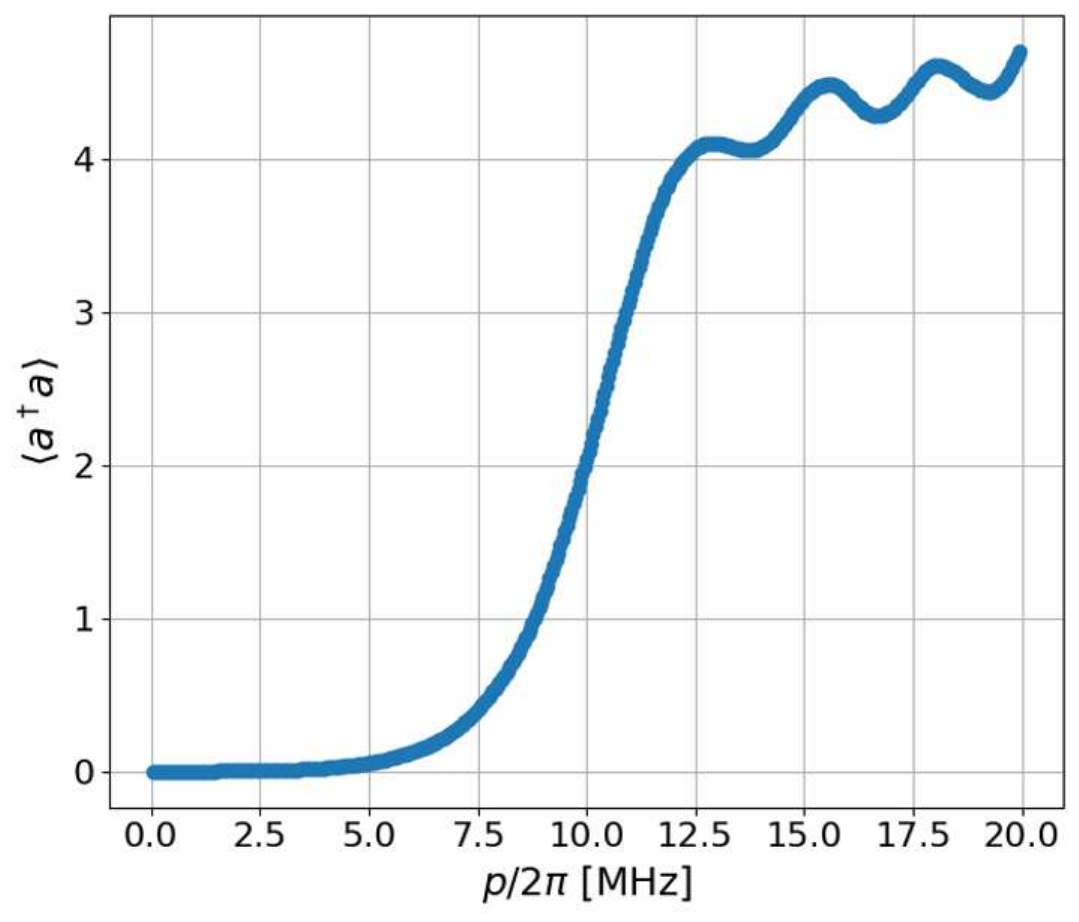}
        \caption{
        Plot at $t=0.2\,\mu$s. The effect of decoherence becomes stronger, and the oscillations weaken.  
        }
        \label{fig:de1}
    \end{subfigure}
    \hfill
    \begin{subfigure}[t]{0.25\textwidth}
        \centering
        \includegraphics[width=\linewidth]{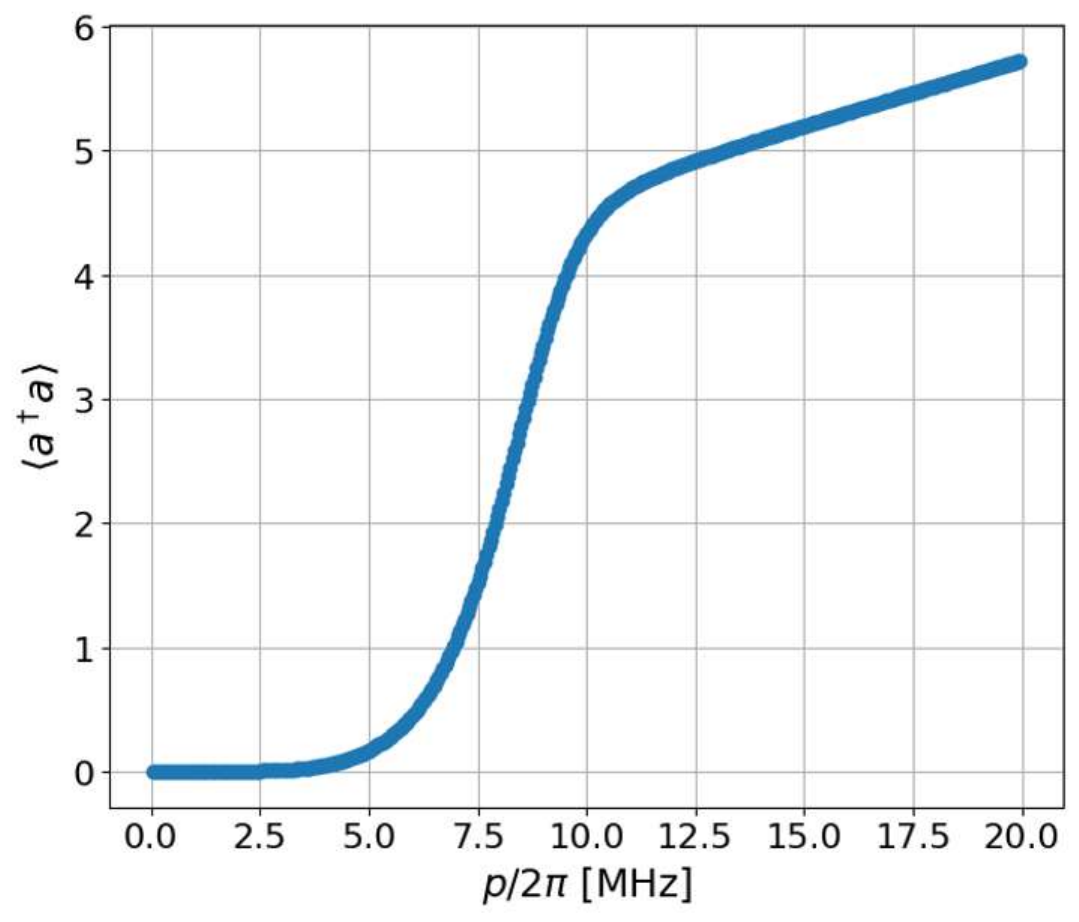}
        \caption{
        Plot at $t=1\,\mu$s. The system approaches a steady state.
        }
        \label{fig04t1}
    \end{subfigure}

    \caption{
    Rabi oscillations under the degeneracy condition between $|0\rangle$ and $|8\rangle$, with the detuning set to $\Delta/2\pi=-65.551$ MHz. 
    The $x$-axis represents the parametric drive $p/2\pi$ [MHz], and the $z$-axis represents the photon number expectation value $\langle a^{\dagger}a\rangle$. 
    The parameters are set to the dissipation rate $\kappa/2\pi=0.73\,$MHz and the nonlinearity $\chi/2\pi=18.729\,$MHz.
 }
    \label{rabi08}
\end{figure}
\begin{figure}[htbp]
    \centering

    \begin{subfigure}[t]{0.32\textwidth}
        \centering
        \includegraphics[width=\linewidth]{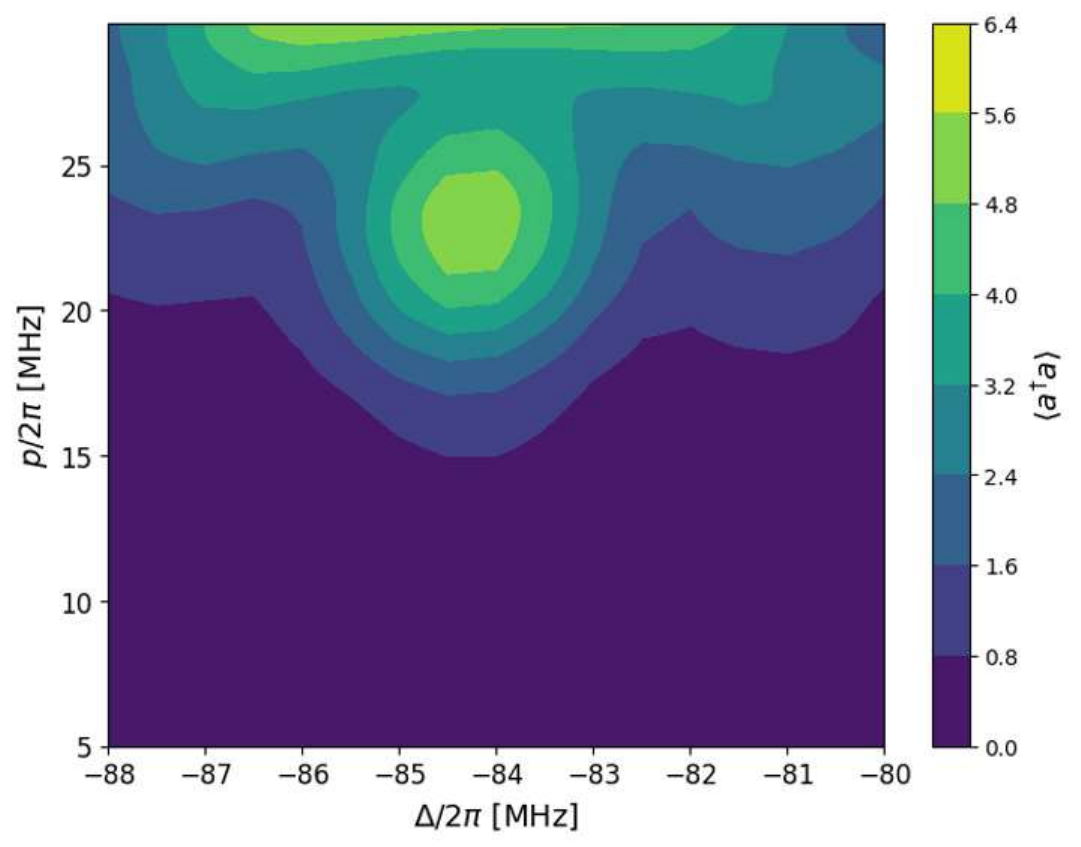}
        \caption{
        Plot at $t=0.05\,\mu$s.    
        }
        \label{fig0_005_3d}
    \end{subfigure}
    \hfill
    \begin{subfigure}[t]{0.32\textwidth}
        \centering
        \includegraphics[width=\linewidth]{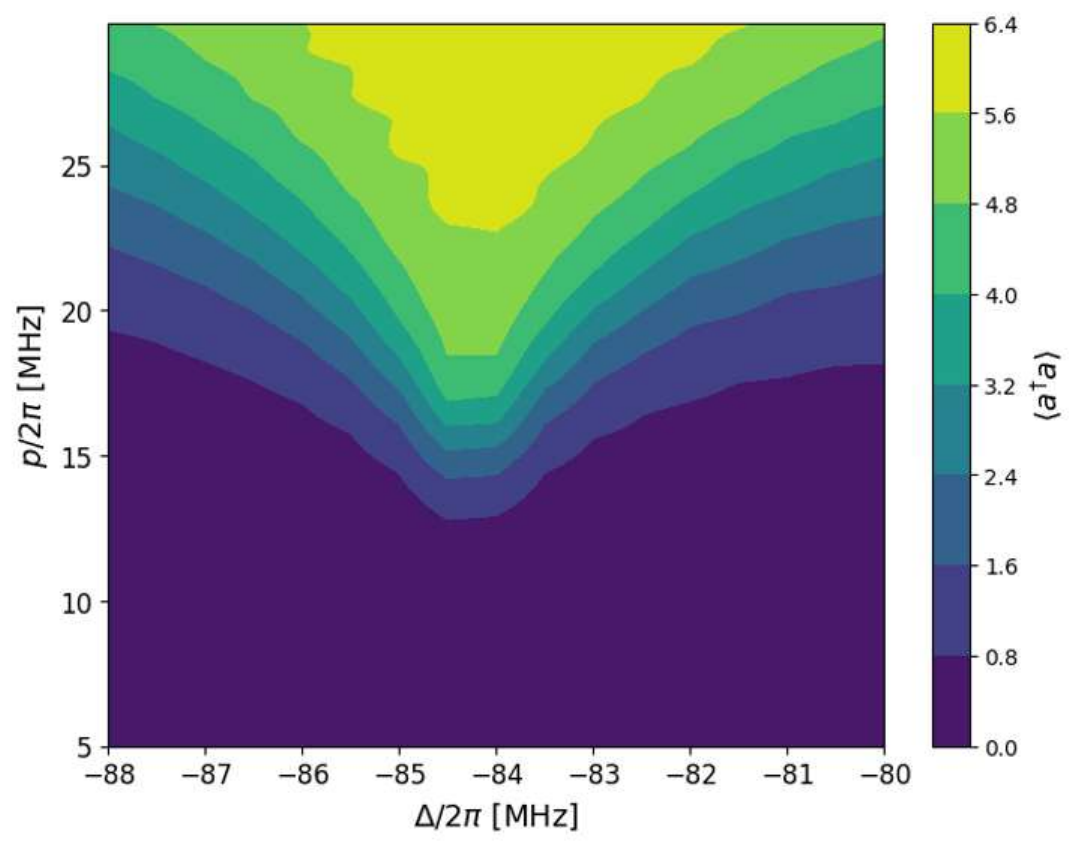}
        \caption{
        Plot at $t=0.2\,\mu$s.  
        }
        \label{fig:deg1}
    \end{subfigure}
    \hfill
    \begin{subfigure}[t]{0.32\textwidth}
        \centering
        \includegraphics[width=\linewidth]{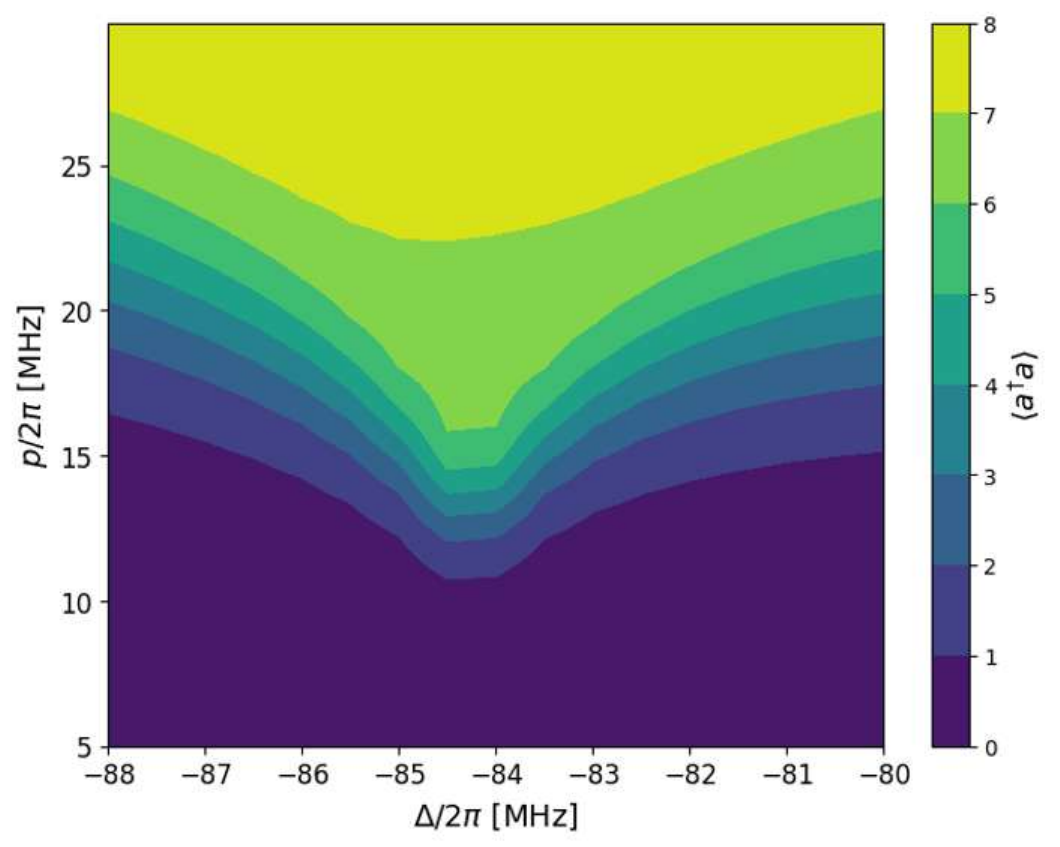}
        \caption{
        Plot at $t=1\,\mu$s. 
        }
        \label{figde06}
    \end{subfigure}

    \caption{
    The master equation was solved, and the photon number expectation value of the KPO was plotted near the degeneracy condition between $|0\rangle$ and $|10\rangle$. 
    The $x$-axis represents the detuning $\Delta/2\pi$ [MHz], the $y$-axis represents the normalized parametric drive $p/\chi$, and the $z$-axis represents the photon number expectation value $\langle a^{\dagger}a\rangle$. 
    The parameters are set to the dissipation rate $\kappa/2\pi=0.73\,$MHz and the nonlinearity $\chi/2\pi=18.729\,$MHz.}
    \label{010_3dim}
\end{figure}
\begin{figure}[htbp]
    \centering

    \begin{subfigure}[t]{0.25\textwidth}
        \centering
        \includegraphics[width=\linewidth]{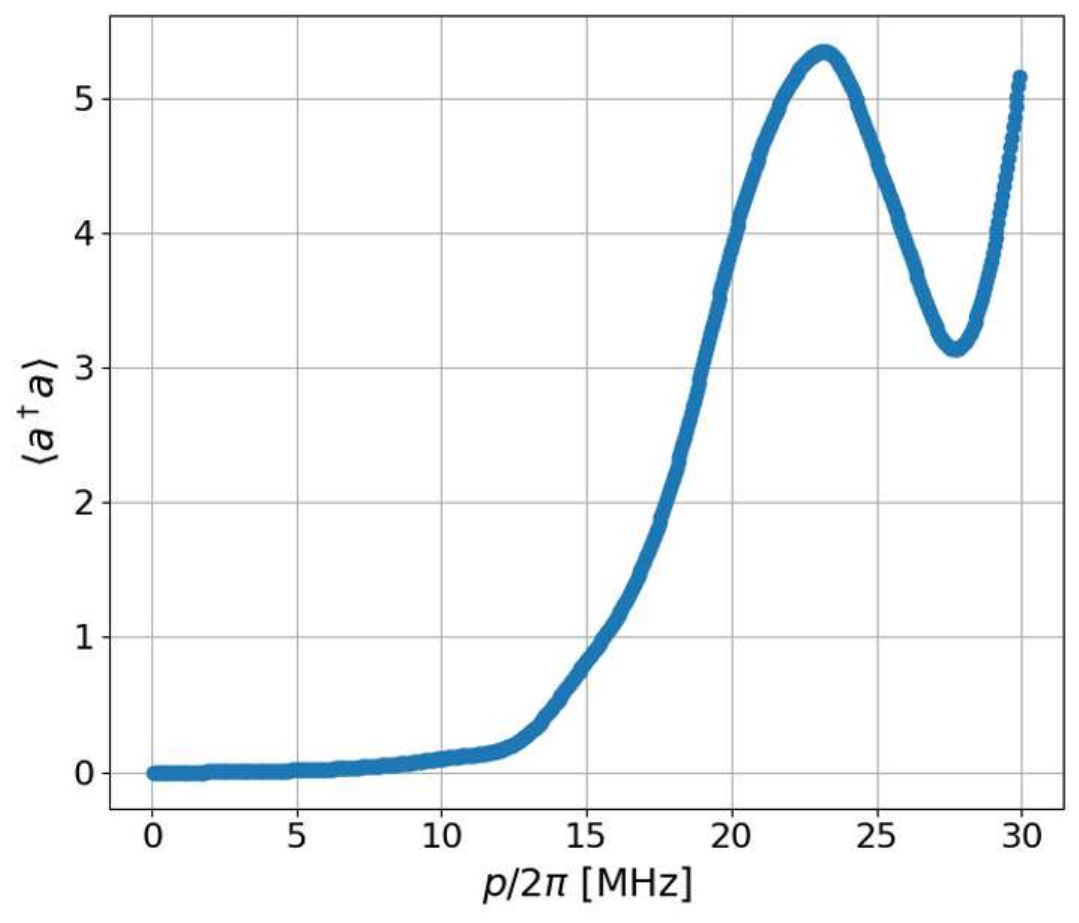}
        \caption{
        Plot at $t=0.05\,\mu$s. Rabi oscillations can be observed.
        }
        \label{rabi010t005}
    \end{subfigure}
    \hfill
    \begin{subfigure}[t]{0.25\textwidth}
        \centering
        \includegraphics[width=\linewidth]{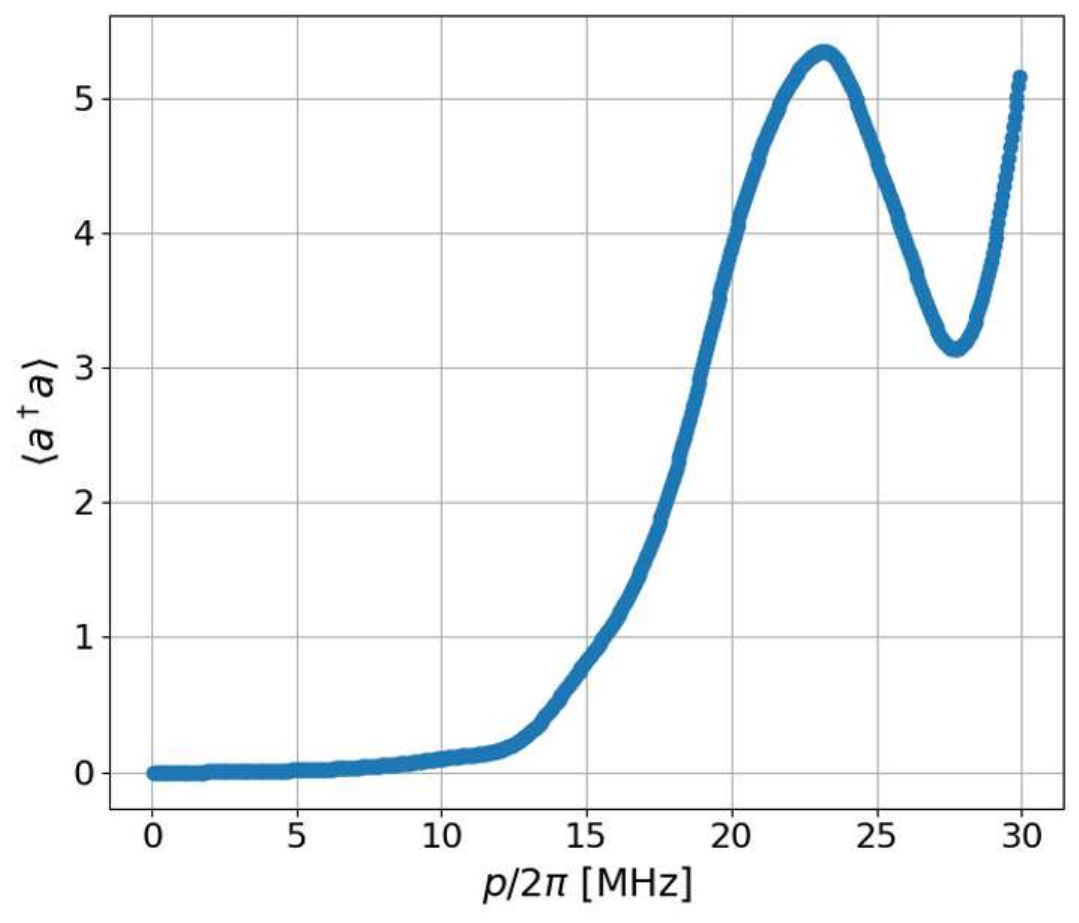}
        \caption{
        Plot at $t=0.2\,\mu$s. The effect of decoherence becomes stronger, and the oscillations weaken.
        }
        \label{rabi010t02}
    \end{subfigure}
    \hfill
    \begin{subfigure}[t]{0.25\textwidth}
        \centering
        \includegraphics[width=\linewidth]{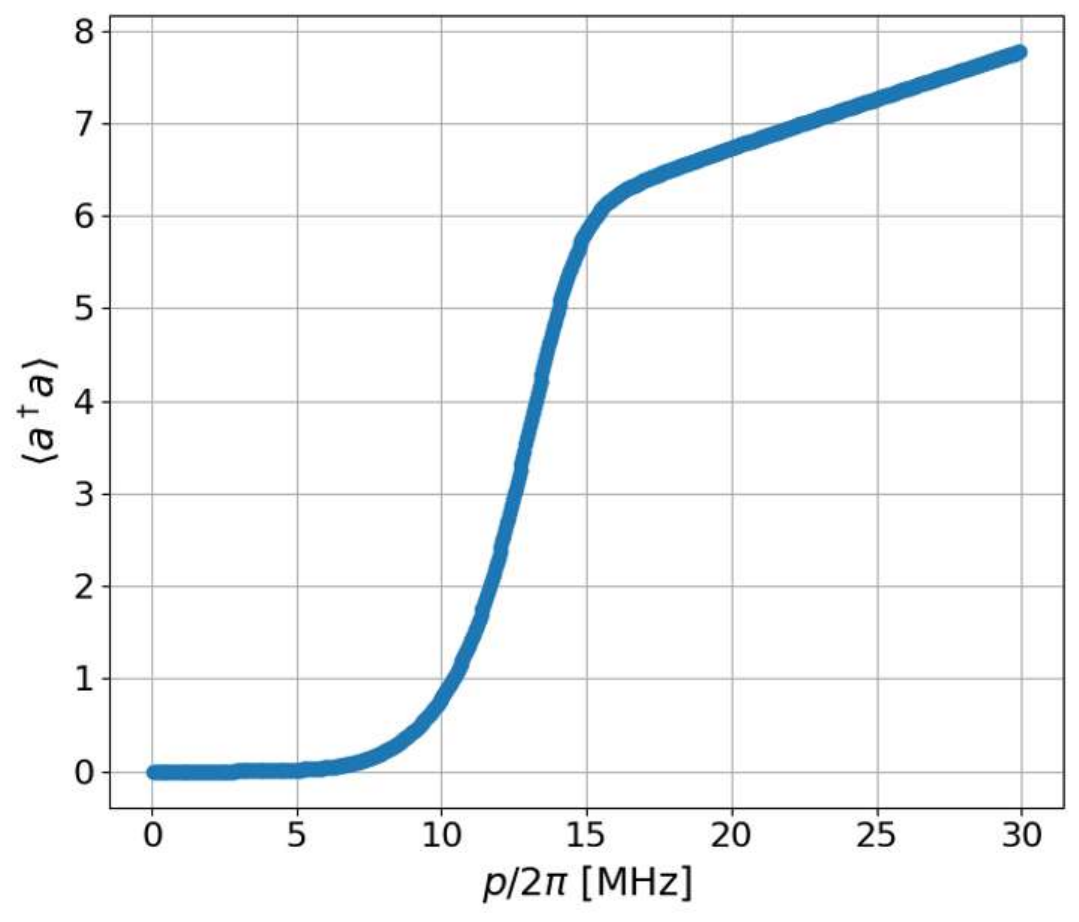}
        \caption{
        Plot at $t=1\,\mu$s. The system approaches a steady state.
        }
        \label{rabi010t1}
    \end{subfigure}

    \caption{
    Rabi oscillations under the degeneracy condition between $|0\rangle$ and $|10\rangle$, where the detuning is set to $\Delta/2\pi=-84.2805$ MHz. 
    The $x$-axis represents the parametric drive $p/2\pi$ [MHz], and the $z$-axis represents the photon number expectation value $\langle a^{\dagger}a\rangle$. 
    The parameters are set to the dissipation rate $\kappa/2\pi=0.73\,$MHz and the nonlinearity $\chi/2\pi=18.729\,$MHz.
    }
    \label{010rabi}
\end{figure}
\begin{figure}[htbp]
    \centering

    \begin{subfigure}[t]{0.32\textwidth}
        \centering
        \includegraphics[width=\linewidth]{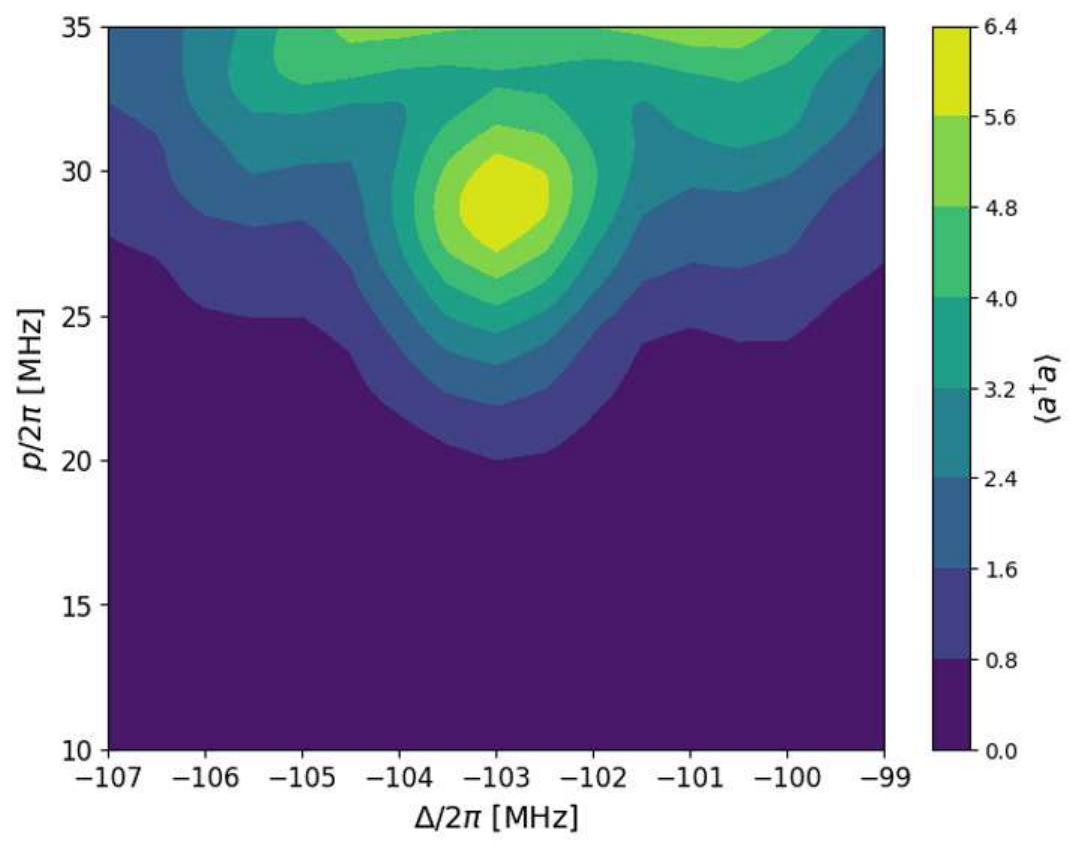}
        \caption{
        Plot at $t=0.05\,\mu$s. 
        }
        \label{012t005_3}
    \end{subfigure}
    \hfill
    \begin{subfigure}[t]{0.32\textwidth}
        \centering
        \includegraphics[width=\linewidth]{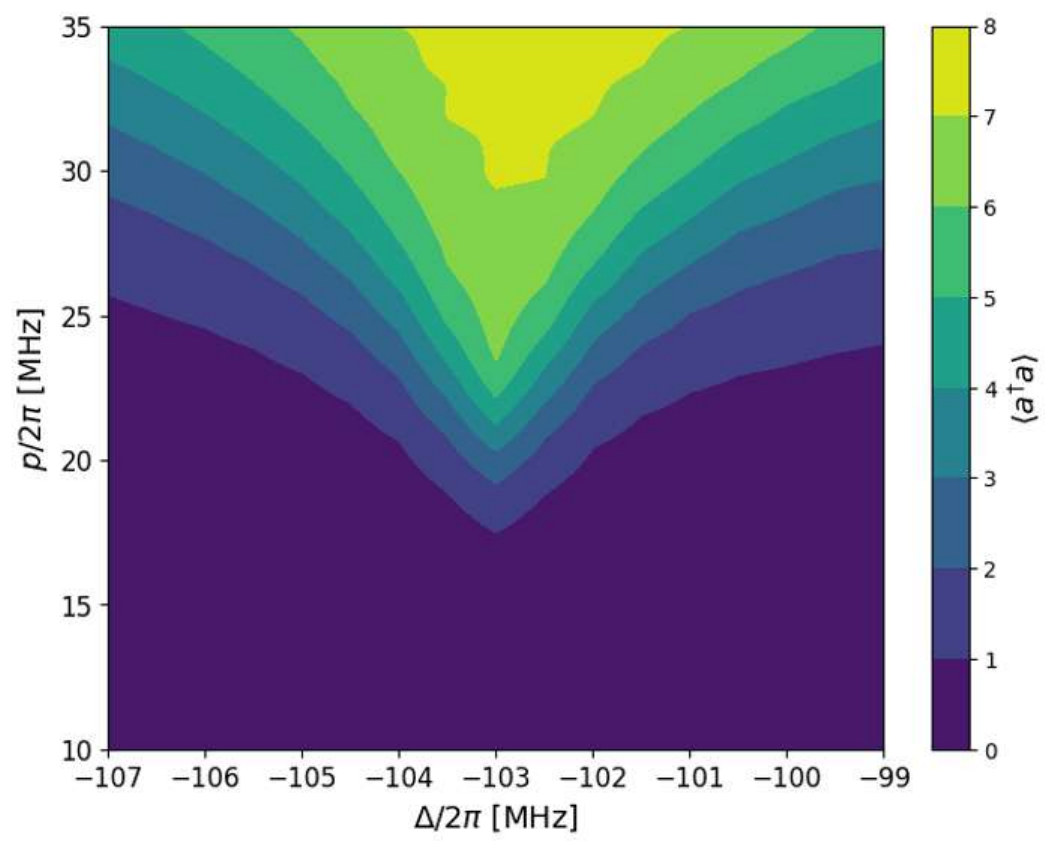}
        \caption{
        Plot at $t=0.2\,\mu$s. 
        }
        \label{012t02_3}
    \end{subfigure}
    \hfill
    \begin{subfigure}[t]{0.32\textwidth}
        \centering
        \includegraphics[width=\linewidth]{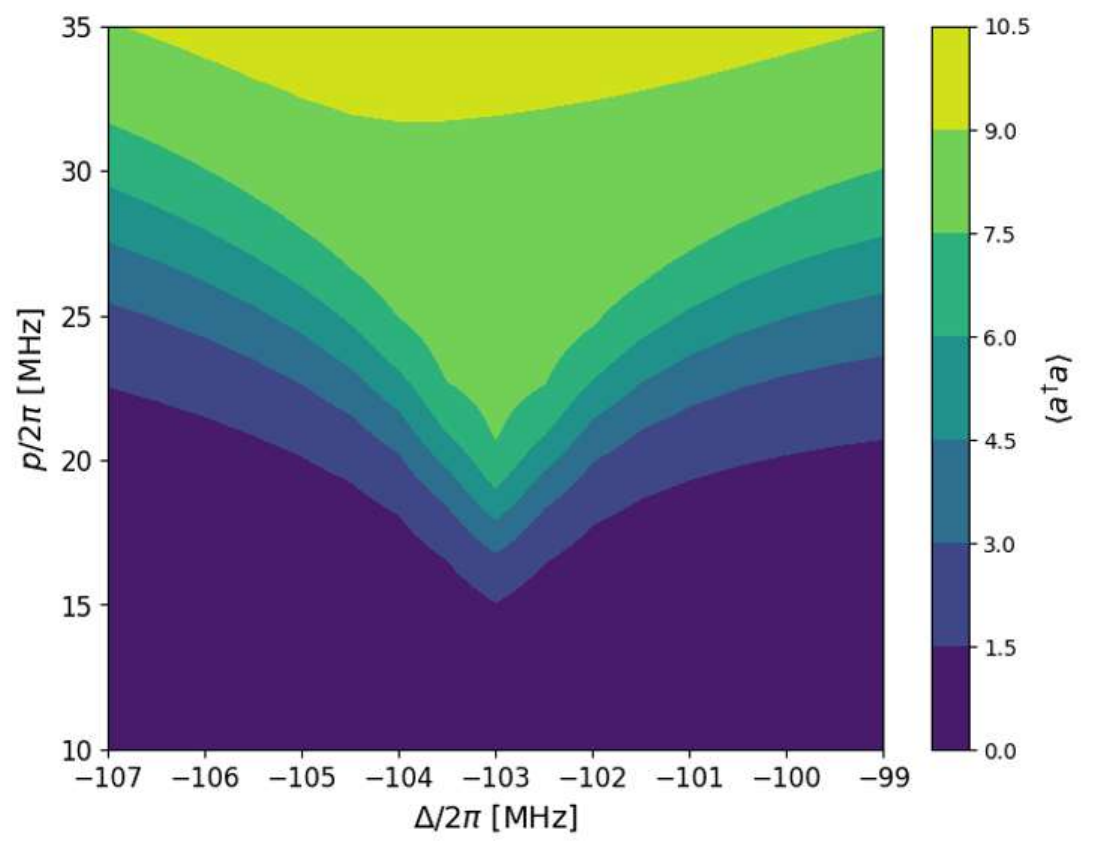}
        \caption{
        Plot at $t=1\,\mu$s. 
        }
        \label{012t1_3}
    \end{subfigure}

    \caption{
    The master equation was solved, and the photon number expectation value of the KPO was plotted near the degeneracy condition between $|0\rangle$ and $|12\rangle$. 
    The $x$-axis represents the detuning $\Delta/2\pi$ [MHz], the $y$-axis represents the normalized parametric drive $p/\chi$, and the $z$-axis represents the photon number expectation value $\langle a^{\dagger}a\rangle$. 
    The parameters are set to the dissipation rate $\kappa/2\pi=0.73\,$MHz and the nonlinearity $\chi/2\pi=18.729\,$MHz.
    }
    \label{012_3dim}
\end{figure}
\begin{figure}[htbp]
    \centering

    \begin{subfigure}[t]{0.25\textwidth}
        \centering
        \includegraphics[width=\linewidth]{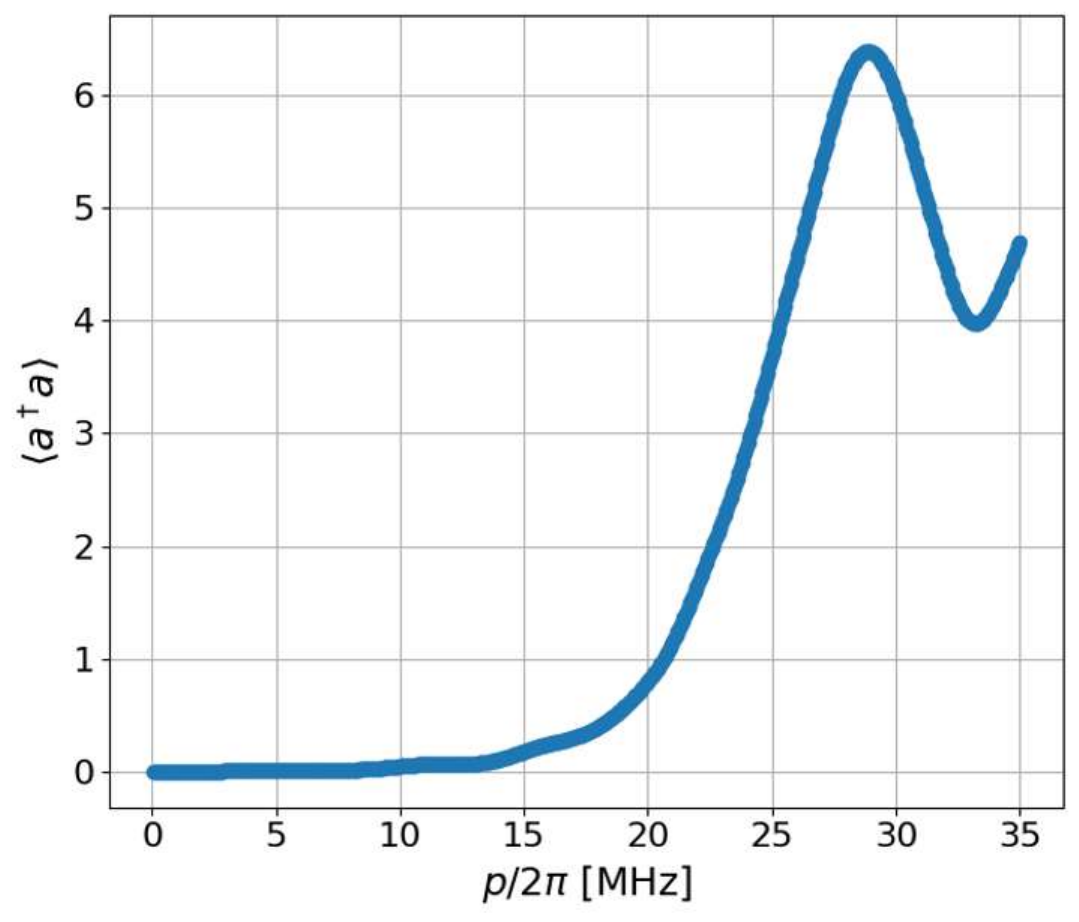}
        \caption{
        Plot at $t=0.05\,\mu$s. Rabi oscillations can be observed.
        }
        \label{rabi012t005}
    \end{subfigure}
    \hfill
    \begin{subfigure}[t]{0.25\textwidth}
        \centering
        \includegraphics[width=\linewidth]{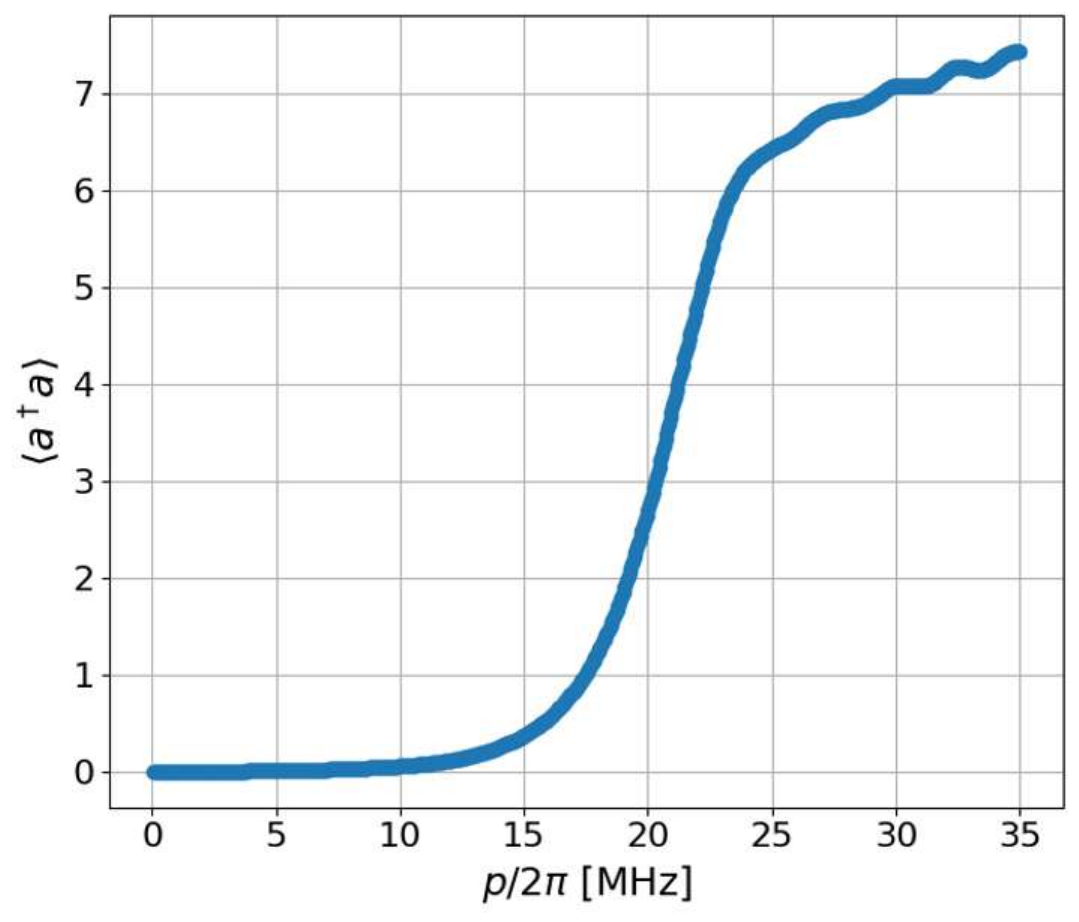}
        \caption{
        Plot at $t=0.2\,\mu$s. The influence of decoherence becomes stronger, and the oscillation amplitude decreases.
        }
        \label{rabi012t02}
    \end{subfigure}
    \hfill
    \begin{subfigure}[t]{0.25\textwidth}
        \centering
        \includegraphics[width=\linewidth]{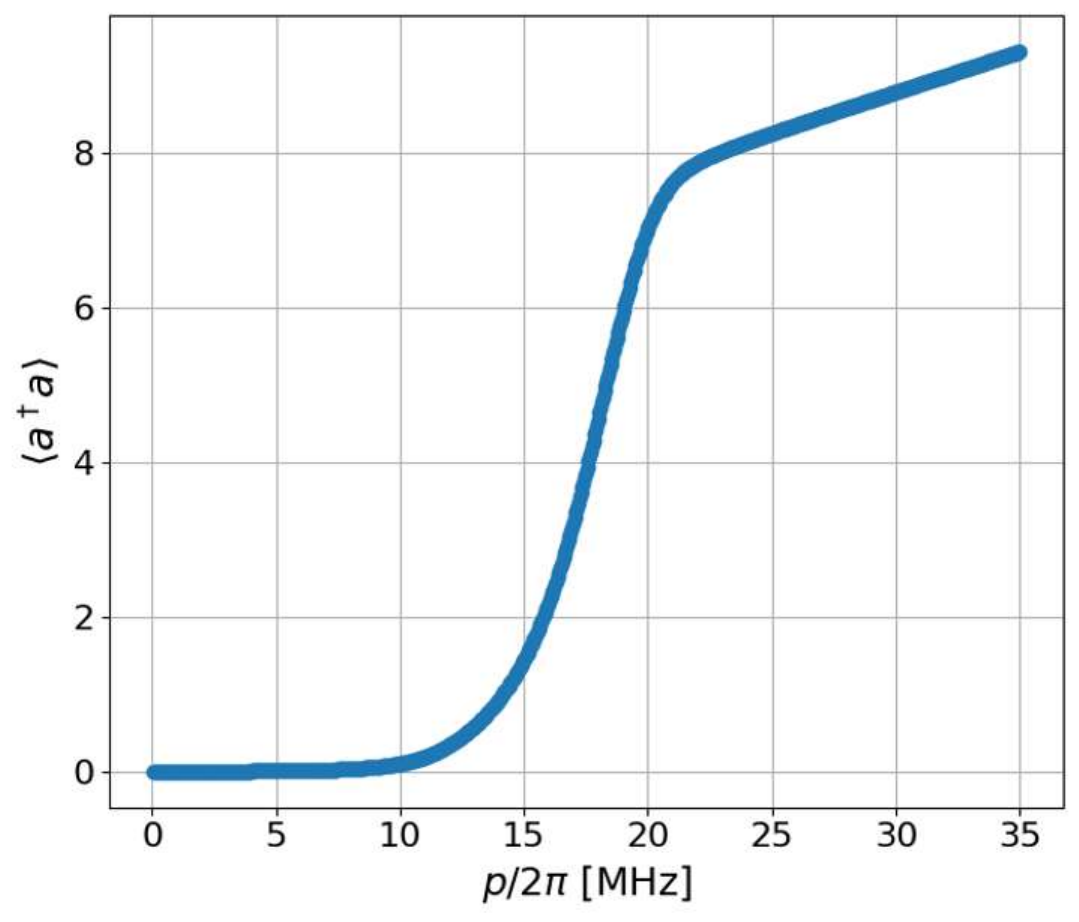}
        \caption{
        Plot at $t=1\,\mu$s. The system approaches the steady state.
        }
        \label{rabi012t1}
    \end{subfigure}

    \caption{
    Time evolution of Rabi oscillations under the degeneracy condition between $|0\rangle$ and $|12\rangle$ states, $\Delta/2\pi=-103.0095$ MHz. The $x$-axis represents the parametric drive amplitude $p/2\pi$ [MHz], while the $z$-axis represents the photon number expectation value $\langle a^{\dagger}a\rangle$. The parameters are set to the dissipation rate $\kappa/2\pi=0.73\,$MHz and the Kerr nonlinearity $\chi/2\pi=18.729\,$MHz.
 }
    \label{012rabi}
\end{figure}
\twocolumngrid
\YM{In this section, we analyze the dynamics of the KPO using numerical simulations implemented with QuTiP~\cite{johansson2012qutip}. In particular, we explain that the phenomenon of PR, theoretically predicted by Bartolo~\cite{bartolo2016exact,roberts2020driven} and \textcolor{black}{also observed in our experiments}, can be understood as coherent oscillations between the ground and excited states and their subsequent decay.}
We describe the time evolution including decoherence by employing the GKSL-type master equation~\eqref{GKSL}, where single-photon loss $a$ is considered as the dissipation channel.

\YM{Figures~\ref{04_3dim} and~\ref{rabi04} show the Rabi oscillations between $|0\rangle$ and $|4\rangle$ under the influence of decoherence, and Figures~\ref{06_3dim} and~\ref{rabi06} show those between $|0\rangle$ and $|6\rangle$. To clearly observe the Rabi oscillations, we set the dissipation rate smaller than the experimental value, namely $\kappa/2\pi=5\times 10^{\tana{-3}}\,$MHz. Subsequently, Figures~\ref{08_3dim} and~\ref{rabi08} present the dynamics between $|0\rangle$ and $|8\rangle$, Figures~\ref{010_3dim} and~\ref{010rabi} for $|0\rangle$ and $|10\rangle$, and Figures~\ref{012_3dim} and~\ref{012rabi} for $|0\rangle$ and $|12\rangle$. Here, we adopt the experimental dissipation rate to enable direct comparison with experiments.
In all cases, when the evolution time $t$ under the parametric drive is much shorter than the inverse of the decay rate, coherence is maintained and oscillations can be observed. As $t$ increases, the oscillations gradually decay, and once the system reaches the steady state, PR is observed.
This demonstrates that the mechanism of PR originates from coherent oscillations induced by higher-order perturbative effects, which decay due to decoherence.
}

\section{Conclusion}
In this study, we theoretically investigated the physical origin of PR in the KPO using both analytical and numerical approaches. 
 PR is a phenomenon where resonance occurs when the detuning is set to $n/2$ times the Kerr nonlinearity, with $n$ being a natural number.
\textcolor{black}{For the analytical treatment based on perturbation theory, we focus on regimes with relatively few photons. In contrast, for large photon numbers, as realized in our experiments, we rely on numerical methods.}
We also compared the theoretical results with experimental data. 
By combining analytical and numerical approaches, we showed that PR proceeds in two stages: a dominant stage of higher-order–induced Rabi oscillations followed by their damping due to decoherence.
\textcolor{black}{Our results deepen the understanding of the properties of the KPO, and as a future direction, we will investigate how these properties can be exploited for applications in quantum information processing.}

\begin{acknowledgements}
This work was supported by the New Energy and Industrial Technology Development Organization (NEDO) under project number JPNP16007.
This work is supported by JST Moonshot (Grant Number JPMJMS226C), CREST (JPMJCR23I5), and Presto JST (JPMJPR245B).
\end{acknowledgements}




\FloatBarrier
\bibliography{ref}

\end{document}